\documentclass[pra,twocolumn,showpacs,preprintnumbers,amsmath,amssymb,superscriptaddress]{revtex4-1}
\usepackage[colorlinks,linkcolor=blue,urlcolor=blue,anchorcolor=blue,citecolor=blue]{hyperref}
\usepackage{amsmath}
\usepackage{graphicx}
\usepackage{dcolumn}
\usepackage{mathrsfs}
\usepackage{amssymb}
\usepackage{amsfonts,multirow}
\usepackage{bm,soul,lineno}

\usepackage{epstopdf}
\usepackage{amsmath}
\usepackage{graphicx}
\usepackage{dcolumn}
\usepackage{verbatim}
\usepackage{mathrsfs}
\usepackage{amssymb}
\usepackage{amsfonts}
\usepackage{bm}

\newif\ifistoreview

\begin{document}
	\draft
	\title{Mixed-State Topology in Non-Hermitian Systems}
	\author{Shou-Bang Yang}
            \affiliation{Fujian Key Laboratory of Quantum Information and Quantum
        		Optics, College of Physics and Information Engineering, Fuzhou University,
        		Fuzhou, Fujian, 350108, China}
        \author{Pei-Rong Han} 
            \affiliation{School of Physics and Mechanical and Electrical Engineering, Longyan University, Longyan, China}
	\author{Wen Ning}
            \affiliation{Fujian Key Laboratory of Quantum Information and Quantum
        		Optics, College of Physics and Information Engineering, Fuzhou University,
        		Fuzhou, Fujian, 350108, China}
	\author{Fan Wu}
            \affiliation{Fujian Key Laboratory of Quantum Information and Quantum
    		Optics, College of Physics and Information Engineering, Fuzhou    University,
    		Fuzhou, Fujian, 350108, China}
	\author{Zhen-Biao Yang} \email{zbyang@fzu.edu.cn}
            \affiliation{Fujian Key Laboratory of Quantum Information and Quantum
		Optics, College of Physics and Information Engineering, Fuzhou University,
		Fuzhou, Fujian, 350108, China}
	\author{Shi-Biao Zheng} 
            \affiliation{Fujian Key Laboratory of Quantum Information and Quantum
		Optics, College of Physics and Information Engineering, Fuzhou University,
		Fuzhou, Fujian, 350108, China}
	
	\vskip0.5cm
	
	\narrowtext
	
	\begin{abstract}
	Non-Hermitian (NH) systems, owing to the existence of exceptional point (or ring and surface), exhibit exotic topological features which are inaccessible in Hermitian systems. While current studies on NH topology has primarily focused on pure states at zero temperature, the topological properties of mixed states remain largely unexplored. In this work, we investigate the mixed-state topology in two-dimensional NH systems using the Uhlmann phase and the thermal Uhlmann-Chern number, both structured via the Uhlmann connection at specific temperatures, revealing distinct topological characteristics compared to those of pure states. Furthermore, we extend our analysis to mixed states in three-dimensional Abelian and four-dimensional non-Abelian NH systems, confirming the existence of the higher-order mixed-state topology. Our study establishes a conceptual and practical pathway for exploring topological phenomena in the mixed-state regime of NH physics.
	\end{abstract}
	
	\maketitle
	\section{Introduction}
		The discovery of topological insulators \cite{RevModPhys.82.3045,RevModPhys.83.1057,shentopological2012,bernevigtopological2013, 
        PhysRevLett.61.2015,PhysRevLett.95.146802,science.1133734,science.1148047,PhysRevLett.98.106803,PhysRevB.76.045302,hsiehtopological2008,RevModPhys.88.035005} has sparked intense interest in uncovering topological nature of quantum materials. 
        This topology is well characterized by the Berry phase  \cite{noauthorquantal1984}, acquired during cyclic adiabatic evolution of a quantum state in momentum or parameter space \cite{jotzu_experimental_2014,roushan_observation_2014}, which is purely geometric in origin. The local curvature in momentum or parameter space, integrated over a closed surface, yields another quantized topological invariant, the first Chern number \cite{jotzu_experimental_2014,roushan_observation_2014,rayobservation2014,noauthorquantised1931,sawadasemi-empirical1974,jackiwdiracs2004}. As a global topology, the Chern number characterizes a system's topological class and gives rise to observable effects like quantized Hall conductance \cite{PhysRevLett.61.2015,PhysRevLett.95.146802,science.1133734,science.1148047}. Recent years have witnessed significant theoretical and experimental progress in probing topological properties across diverse physical systems, including superconducting circuits \cite{PhysRevLett.113.050402,roushan_observation_2014,PhysRevLett.126.017702}, atomic systems \cite{PhysRevA.98.013603,PhysRevA.89.013627,PhysRevLett.125.217202}, and photonic systems \cite{PhysRevA.98.033830,PhysRevA.97.031801,PhysRevLett.119.183901}. Howerver, most of these studies isolate quantum systems from their surrounding environment to minimize decoherence effects. 
		
		Non-Hermitian (NH) systems, encompassing both unitary and dissipative (gain-and-loss) physics, exhibit distinctive features absent in Hermitian cases, including spectral transitions \cite{PhysRevLett.86.787,PhysRevLett.104.153601,gao_observation_2015,zhang_observation_2017}, symmetry \cite{PhysRevLett.80.5243,PhysRevLett.89.270401,ozdemir_paritytime_2019,PhysRevLett.103.093902,feng_single-mode_2014,hodaei_parity-timesymmetric_2014,PhysRevLett.124.070402,PhysRevLett.126.170506,ren_chiral_2022}, dynamical effects \cite{PhysRevX.8.021066,doppler_dynamically_2016,xu_topological_2016,yoon_time-asymmetric_2018}, entanglement transitions \cite{PhysRevLett.131.260201}, sensitivity enhancement \cite{chen_exceptional_2017,hodaei_enhanced_2017} and NH topology \cite{s11433_025_2851_8,yang2025hyperspherelikenonabelianyangmonopole,bergholtz_exceptional_2021,ding_non-hermitian_2022}. The rich phenomenology of NH systems is closely tied to exceptional points (EPs), where both eigenenergies and eigenstates coalesce \cite{miri_exceptional_2019,bergholtz_exceptional_2021,ding_non-hermitian_2022}. Furthermore, the discovery of the extention of EPs, such as exceptional rings (ERs) \cite{PhysRevLett.118.045701,PhysRevB.99.121101,PhysRevLett.127.196801,PhysRevB.104.L161117,PhysRevResearch.2.043268,PhysRevB.100.245205,PhysRevLett.129.084301,zhen_spawning_2015,cerjan_experimental_2019} and exceptional surfaces (ESs) \cite{tang_realization_2023,zhou_exceptional_2019,PhysRevLett.123.237202}- has greatly expanded the scope of NH physics.
		
		On the other hand, when the thermal noise effect from the environment is considered, an NH quantum system is described as a mixed state at finite temperature \cite{sakurai_modern_2011}. Recently, mixed-state physics and applications have attracted extensive interest, covering topics such as protected symmetry \cite{Coser2019classificationof,PhysRevX.15.011069,xue2024tensornetworkformulationsymmetry,PRXQuantum.6.010348,PhysRevB.111.115141,shah2024instabilitysteadystatemixedstatesymmetryprotected,PRXQuantum.6.020333,PhysRevB.110.165160,PhysRevX.15.021060}, quantum error correction \cite{PhysRevX.14.031044,PhysRevB.110.085158,PhysRevB.111.125106}, quantum encoding \cite{PhysRevLett.134.070403,hauser2024informationdynamicsdecoheredquantum,negari2025spacetimemarkovlengthdiagnostic}, topology \cite{hchr_rqq9,bao2023mixedstatetopologicalordererrorfield,PRXQuantum.5.020343,PhysRevX.15.021062,PRXQuantum.4.030318,PhysRevLett.132.170602,PhysRevB.110.125152,PRXQuantum.6.010314,PhysRevB.109.035146,PhysRevB.110.125145,PRXQuantum.6.010313,PRXQuantum.6.010315,fy9r-hpcw,kim2025persistenttopologicalnegativityhightemperature,Wang_2025} and spontaneous symmetry breaking \cite{PRXQuantum.5.030310,PRXQuantum.6.010344,PhysRevB.110.155150,gu2024spontaneoussymmetrybreakingopen,PhysRevB.111.115137,kim2024errorthresholdsykcodes,PhysRevB.111.125147,9kmh-gjf8}.
        The topology of mixed states can be probed via the Uhlmann connection, a geometric extension of the Berry connection to density matrices \cite{uhlmann_parallel_1986,uhlmann_gauge_1991,uhlmann_density_1993}. The Uhlmann phase, accumulated during cyclic evolution of the density matrix in a Uhlmann process, serves as a finite-temperature topological indicator \cite{PhysRevLett.85.2845,PhysRevLett.91.090405,PhysRevA.75.032106,zhu_experimental_2011,PhysRevB.91.165140,andersson_geometric_2016,mera_boltzmanngibbs_2017,PhysRevLett.119.015702}. Numerous theoretical studies on the Uhlmann phase and mixed-state topology have been proposed \cite{PhysRevLett.112.130401,PhysRevB.106.024310,PhysRevB.97.235141,prq8-c9ns,kartik_mixed_2023,PhysRevB.105.085418,PhysRevResearch.5.023004,PhysRevX.8.011035,PhysRevB.107.165415,liu_uhlmann_2022,carollo_uhlmann_2018}, and the experimental measurement of the Uhlmann phase has also been demonstrated \cite{viyuela_observation_2018}. Nevertheless, their exploration in NH systems remains largely unexplored.
        
		We first construct a two-dimensional (2D) NH system featuring an exceptional ring (ER), arising from unitary dynamics combined with both dissipative and thermal environmental effects. The topology of the ER is characterized by
        the Uhlmann phase and the thermal Uhlmann-Chern number \cite{PhysRevB.97.235141,prq8-c9ns}, both of which reveal exceptional features distinct from the those in open systems with pure NH effects. We then extend our investigation of such exceptional topology to higher dimensions. In a 3D NH system \cite{s11433_025_2851_8}, we introduce a thermal Dixmier-Douady (DD) invariant to characterize its finite-temperature topology. For a 4D non-Abelian NH system \cite{yang2025hyperspherelikenonabelianyangmonopole}, we analyze the Uhlmann phase and the second thermal Uhlmann-Chern number, both of which demonstrate higher-order topological features at finite temperatures.
		Our work advances the understanding of mixed-state topology in higher-dimensional systems by unifying NH physics and quantum geometry.
		
	\section{Mixed-State Topology of the 2D NH system}
	\subsection{The Uhlmann phase}
	We consider a generic two-level system with particle gain and loss, the Hamiltonian is (setting $\hbar=1$)
	\begin{eqnarray}
		H=\sum_{\nu={x,y,z}}q_\nu\sigma_\nu+i\gamma\sigma_z,
		\label{eq1}
	\end{eqnarray}
	where $\sigma_\nu$ are Pauli matrices, $q_{\nu}$ the corresponding control parameters and $\gamma$ the gain-loss rate.
      The eigenenergies of Eq. (\ref{eq1}) are $E=\pm\sqrt{\Omega^2-\gamma^2}$, with $\Omega=\sqrt{q_x^2+q_y^2+q_z^2}$. When $\Omega=\gamma$, the two eigenenergies coalesce, and the exceptional point (EP), originally located at the center of the Bloch sphere ($\gamma = 0$), expands into an ER of radius $\gamma$ in the $\{q_x,q_y\}$ plane (taking $q_z = 0$), giving rise to intriguing topological properties.
	
	We now focus on mixed-state topology, which can be probed via the Uhlmann phase. To this end, we construct a parameter loop in the $\{q_x,q_z\}$ plane to encircle the ER. The controlled Hamiltonian in (\ref{eq1}) is parameterized as $\{q_x,q_y,q_z\}=\{r\sin\theta+d,0,r\cos\theta\}$,
    where $d = 5\gamma/2$ denotes the displacement of the loop center along the ${q_x}$ direction and $r = 2\gamma$ is the loop radius. The eigenenergies and eigenstates are denoted by $E_{1,2}$ and $|u_{1,2}\rangle$, respectively. The corresponding normalized left eigenstates $\langle u^L_{1,2}|$ satisfy $\langle u^L_n|H_2=\langle u^L_n|E_n$ and $\langle u^L_m|u_n\rangle=\delta_{mn}$. At finite temperature $T$, the mixed-state density matrix is expressed as
	\begin{align}
		\rho=\sum_{n=1,2}P_n|u_n\rangle\langle u^L_n|,
		\label{eq4}
	\end{align}
	 with Boltzmann weights $P_n={e^{-E_n/T}}/{Z}$ and Z the partition function.
	As $\theta$ evolves from 0 to $4\pi$, the eigenenergies traverse both sides of the Möbius-like energy band, while the mixed-state trajectory encircles the ER twice in parameter space. The Uhlmann connection is given by 
\begin{figure}[t!] 
		\centering
		\includegraphics[width=3.4in]{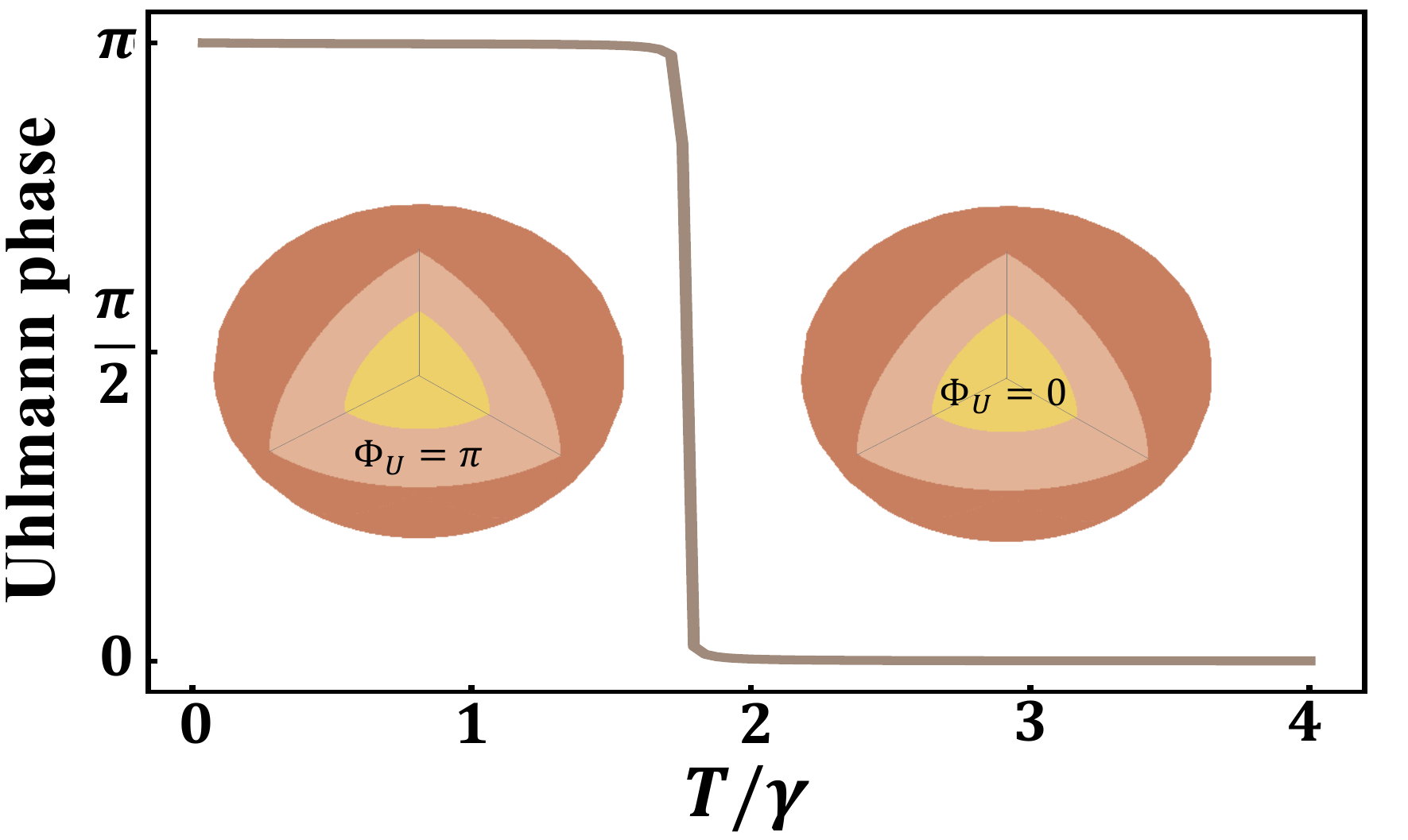}
		\caption{The Uhlmann phase $\Phi_U$ as a function of temperature $T$, which scales with $\gamma$.}
		\label{Fig1}
	\end{figure}%
	\begin{align}
		A_U^\theta=\sum_{m,n}^{1,2}\frac{|u_m(\theta)\rangle\langle u^L_m(\theta)|\left[\partial_\theta\sqrt{\rho_\theta},\sqrt{\rho_\theta}\right]|u_n(\theta)\rangle\langle u^L_n(\theta)|}{P_m(\theta)+P_n(\theta)}\textrm{d}\theta. \nonumber\\
		\label{eq5}
	\end{align}
	Under the parallel transport condition and using Eq.~(\ref{eq4}), this reduces to
	\begin{widetext}
	\begin{eqnarray}
		A_U^\theta&=&\frac{1}{2}\left[f(T)-f^2(T)\right]\left(|u_1\rangle\langle u_2^L|\langle\partial_\theta u_1^L|u_2\rangle- |u_2\rangle\langle u_1^L|\langle u_2^L|\partial_\theta u_1\rangle\right)\nonumber\\
		&&+\frac{1}{2}\left[3f(T)-f^2(T)\right]\left(|u_1\rangle\langle u_2^L|\langle u_1^L|\partial_\theta u_2\rangle- |u_2\rangle\langle u_1^L|\langle \partial_\theta u_2^L|u_1\rangle\right)\nonumber\\
		&=&f(T)\left(|u_1\rangle\langle u_2^L|\langle\partial_\theta u_1^L|u_2\rangle- |u_2\rangle\langle u_1^L|\langle u_2^L|\partial_\theta u_1\rangle\right),
		\label{eq6}
	\end{eqnarray}
	\end{widetext}
	where $f(T)=\left(1-\textrm{sech}{\frac{|E_1|}{T}}\right)$. The Uhlmann phase, determined by the holonomy of $A_U^\theta$, is defined via the mismatch between the initial and final points:
	\begin{figure}[t!] 
		\centering
		\includegraphics[width=3.4in]{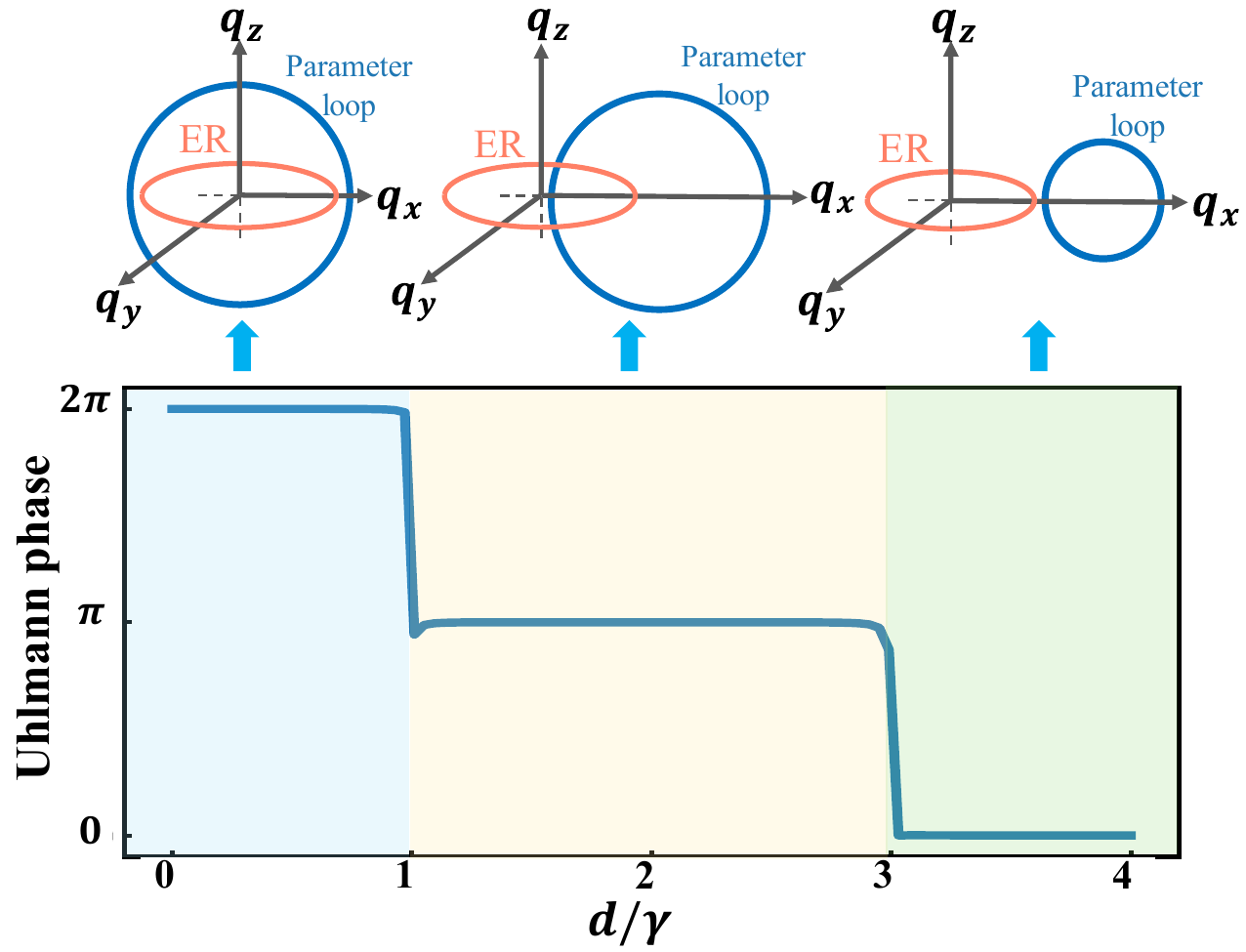}
		\caption{The Uhlmann phase $\Phi_U$ as a function of the parameter loop displacement $d$, which scales with $\gamma$. $\Phi_U$ accumulates 2$\pi$, $\pi$, and 0 after one cycle with two loops of mixed-state evolution when the parameter loop encloses two EPs, encircles the ER while enclosing one EP and disengages from the ER, respectively.}
		\label{Fig2}
	\end{figure}%
	\begin{eqnarray}
		\Phi_U=\textrm{arg Tr}(\rho_0 e^{\int A_U^\theta \textrm{d}\theta}),
	\end{eqnarray}
    with $\rho_0$ the density matrix at $\theta=0$.
	Numerical simulations of $\Phi_U$ are shown in Fig.~\ref{Fig1}. For $T/\gamma<1.95$, the phase accumulates $\pi$ after two looping cycles (2$\mathcal{L}$), around the ER. In contrast, $\Phi_U$ vanishes for $T/\gamma >1.95$, with $T/\gamma =1.95$ marking the critical point of a topological transition.
	This temperature-dependent transition, absent in pure-states NH systems, highlights the interplay between thermal fluctuations and NH topology.
	
	To further investigate this feature, we fix $T/\gamma=0.5$ and gradually displace the parameter loop by tuning $d$ from $0$ to $4\gamma$, as shown in Fig.~\ref{Fig2}. For $d<\gamma$, the parameter loop lies outside the ER and encircles the two EPs on the $\{q_x,q_z\}$ plane. In this case, the Uhlmann phase $\Phi_U$ behaves similarly to that in Hermitian systems: it accumulates a $2\pi$ after the mixed state evolves through one cycle consisting of two loops ($2\mathcal{L}$). When $d$ increases to $d=\gamma$, the parameter loop crosses the ER. At this boundary ($d=\gamma$), $\Phi_U$ undergoes a transition from $2\pi$ to $\pi$, making a first topological transition. As $d$ increases further, the loop crosses the ER and becomes intertwined with it. When the loop encircles only one EP on the $\{q_x,q_z\}$ plane, $\Phi_U$ remains $\pi$ until $d$ increases up to $3\gamma$. At this second boundary ($d = 3\gamma$), $\Phi_U$ undergoes another topological transition, from $\pi$ to $0$. For even larger $d$, the loop moves far away from the ER and no longer encloses it, during which $\Phi_U$ stays at zero. The twofold topological transition stands in stark contrast to the Uhlmann phase behavior observed in Hermitian systems. 
    
	\subsection{The Thermal Uhlmann-Chern number}
	We now investigate the global topology characterized by the thermal Uhlmann-Chern number when the parameter space extends to a 3D sphere at finite temperatures. The Hamiltonian in Eq.~(\ref{eq1}) is parameterized by 
	$\{q_x,q_y,q_z\}=R\{\sin\theta\cos\phi,\sin\theta\sin\phi,\cos\theta\}$, where $R$ denotes the sphere radius. The eigenenergies are $E_{1,2}=\pm\sqrt{R^2-\gamma^2+2i\gamma R\cos{\theta}}$, and the corresponding right eigenstates are $|u_{1,2}\rangle=
	 	\left(
	 		R\sin{\theta}e^{-i\phi},
	 		E_{1,2}-R\cos{\theta}-i\gamma
	 	\right)/N_{1,2}$,
	\begin{figure}[b!] 
		\centering
		\includegraphics[width=3.4in]{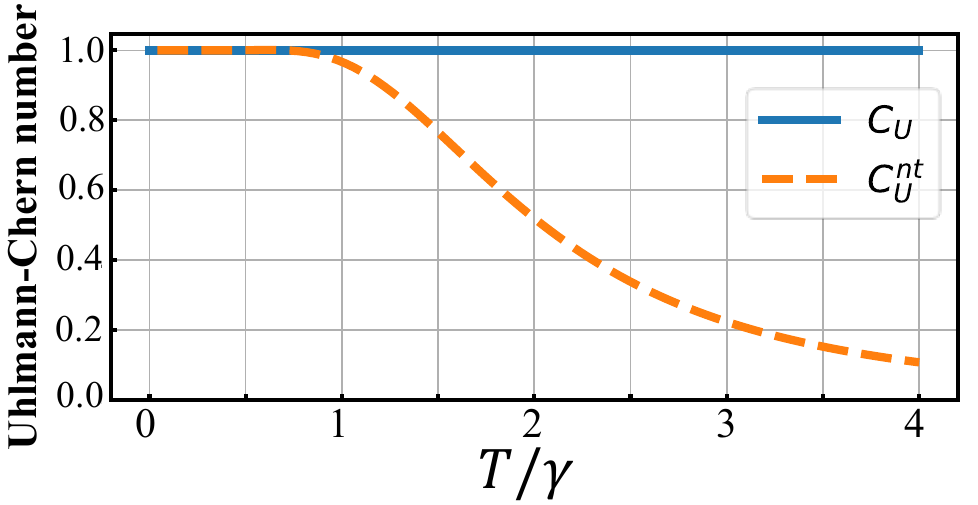}
		\caption{The thermal Uhlmann-Chern number $C_U$ (blue solid line) and NT thermal Uhlmann-Chern number $C_U^{nt}$ (yellow dashed line) as functions of temperature $T$, which scales with $\gamma$.}
		\label{Fig3}
	\end{figure}%
	where $N_{1,2}$ is the normalization constants. The left eigenstates $\langle u_{1,2}^L|$ follow from the biorthogonal condition $\langle u_m^L|u_n\rangle=\delta_{mn}$ \cite{PhysRevLett.131.260201}. From Eq.~(\ref{eq6}), we obtain the Uhlmann connections $A_U^\theta$ and $A_U^\phi$ at temperature $T$.
	The thermal Uhlmann-Chern number is then determined by the Uhlmann curvature \cite{PhysRevB.97.235141}. Importantly, unlike the Berry connection, the Uhlmann connections derived from the mixed-state density matrix are inherently matrix-valued. The Uhlmann curvature is defined as \cite{uhlmann_parallel_1986}
	\begin{eqnarray}
		F_U=\textrm{d}A_U+A_U\wedge A_U,
		\label{eq10}
	\end{eqnarray}
    and the thermal Uhlmann-Chern number is given by
    \begin{eqnarray}
		C_U = \frac{i}{2\pi}\int \lambda(E,T)\textrm{Tr}(\rho F_U), 1/\lambda(E,T)=\tanh^3{\frac{E}{T}}.
		\label{eq14}
	\end{eqnarray}
		Substituting the Uhlmann connection from Eq.~(\ref{eq6}) into Eq.~(\ref{eq10}) and evaluating the trace with the density matrix yields
    \begin{widetext}
	\begin{eqnarray}
		\textrm{Tr}\left(\rho \partial_\theta A_U^\phi\right)&=&-\textrm{Tr}\left(\rho \partial_\phi A_U^\theta\right)=\tanh{\frac{E}{T}}f(T)\frac{iR^2\sin{\theta}^2}{|N|^2}\left(\langle\partial_\theta u_-^L|u_+\rangle-\langle u_+^L|\partial_\theta u_-\rangle\right),   \nonumber\\
		\textrm{Tr}\left(\rho\left[A_U^\theta,A_U^\phi\right] \right)&=&\tanh{\frac{E}{T}}f^2(T)\frac{iR^2\sin{\theta}^2}{|N|^2}\left(\langle\partial_\theta u_-^L|u_+\rangle-\langle u_+^L|\partial_\theta u_-\rangle\right).
	\end{eqnarray}
    \end{widetext}
	Inserting the density matrix into the definition of the Chern character to construct the form
	\begin{equation}
		\textrm{Tr}(\rho F_U) = \tanh^3{\frac{E}{T}}\frac{iR^2\sin{\theta}^2}{|N|^2}\left(\langle\partial_\theta u_-^L|u_+\rangle-\langle u_+^L|\partial_\theta u_-\rangle\right),\\
	\end{equation}
    and plugging this into Eq. (\ref{eq14}), leads to $C_U=1$. Additionally, in the presence of thermal fluctuations, a nontopological (NT) thermal Uhlmann-Chern number can be defined, that is
	\begin{eqnarray}
		C_U^{nt}=\frac{i}{2\pi}\int\textrm{Tr}(\rho F_U),
	\end{eqnarray}
	whose temperature-dependent feature is shown in Fig.~\ref{Fig3}. It clearly shows that the NT thermal Uhlmann-Chern number $C_U^{nt}$ drops gradually with the increase of the temperature $T$. This is contrast to the case of pure states in NH systems, where the Chern number remains $1$.
	\begin{figure}[t!] 
		\centering
		\includegraphics[width=3.4in]{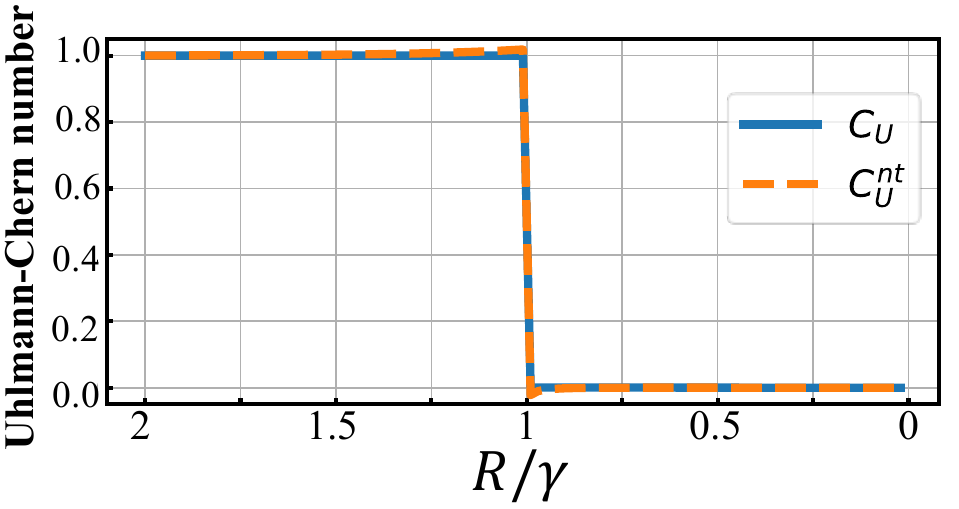}
		\caption{The thermal Uhlmann-Chern number $C_U$ (blue solid line) and the NT thermal Uhlmann-Chern number $C_U^{nt}$ (yellow dashed line) as functions of the parameter sphere radius R, which scales with $\gamma$.}
		\label{Fig4}
	\end{figure}%
	
    We then show the topological transition of the thermal Uhlmann-Chern number $C_U$ and the NT thermal Uhlmann-Chern number $C_U^{nt}$ with respect to the radius $R$ for the fixed temperature $T/\gamma=0.5$.
    As shown in Fig.~\ref{Fig4}, both $C_U$ and $C_U^{nt}$ keep $1$ when $R/\gamma>1$, and abruptly vanish to 0 when crossing the critical point $R/\gamma=1$ and finally keep $0$ for $R/\gamma<1$. Both $C_U$ and $C_U^{nt}$ exhibit a topological transition from the topological phase to the normal phase at the critical point $R/\gamma=1$. The phenomenon is precluded in the Hermitian case due to the spectral properties and the orthogonality of eigenstates. The intriguing feature exhibited in variation of both $C_U$ and $C_U^{nt}$ versus $T$ and $R$ demonstrate that the topological character is jointly influenced by thermal noise and dissipation, with its effect suppressed as the system-environmental coupling strengthens. The interplay of the mixed-state geometry and the NH-degeneracy advances the exotic topology inherent in such a thermal system-environment coupling mechanism.   
	
		\section{Mixed-State Topology of the 3D NH system}
    For a three-level system, the NH Hamiltonian is modeled as
    \begin{eqnarray}
        H_3=\vec{q}\cdot\vec{\Lambda}+i\gamma\Lambda_8,
        \label{eq15}
    \end{eqnarray}
    where $\vec{q}=\{\Omega_1\cos{\phi_1},\Omega_1\sin{\phi_1},\Omega_2\cos{\phi_2},
    \Omega_2\sin{\phi_2}\}$ defines the 4D parameter space, $\vec{\Lambda}=\{\Lambda_1,\Lambda_2,\Lambda_6,\Lambda_7\}$ are the $3\times3$ Gell-Mann matrices \cite{PhysRev.125.1067}, satisfying $[\lambda_j,\lambda_k]=if^{jkl}\lambda_l$. The term $i\gamma\Lambda_8$ introduces the non-Hermiticity.
    For $|\Omega_1|=\gamma/3$ and $|\Omega_2|=2\sqrt{2}\gamma/3$, an exceptional surface (ES) composed of third-order EPs emerges in the 4D parameter space, where the eigenenergies are threefold degenerate and the eigenstates exhibit a three-order exceptional topological transition.
    We then construction the parameters as $\Omega_1=R\cos\alpha$ and $\Omega_2=R\sin\alpha$,
    the topological properties of the pure-state case can be characterized by the Dixmier-Douady (DD) invariant, defined as \cite{PhysRevD.31.1921,chen_synthetic_2022},
		$\mathcal{DD}=\frac{1}{2\pi^2}\int_{S^3}\mathcal{M}_{\alpha\phi_1\phi_2}\textrm{d}\alpha\wedge \textrm{d}\phi_1\wedge \textrm{d}\phi_2,$
    where $\mathcal{M}_{\alpha\phi_1\phi_2}$ is the three-form Berry curvature, related to the quantum metric or the two-form curvature by $\mathcal{M}_{\alpha\phi_1\phi_2}=\epsilon_{\alpha\phi_1\phi_2}\left[4\sqrt{\textrm{det}(\mathcal{G}_{\alpha\phi_1\phi_2})}\right]
		=-\frac{1}{2}\left(\mathcal{F}_{\alpha\phi_1}+\mathcal{F}_{\phi_2\alpha}\right)$.
    In NH systems, the quantum metric tensor and Berry curvature correspond to the real and imaginary parts of the quantum geometric tensor \cite{s11433_025_2851_8} $\chi_{\mu\nu}=\mathcal{G}_{\mu\nu}+i\mathcal{F}_{\mu\nu}=\sum_{n\neq m}\frac{|\langle u_{m}^{L}|\partial_\mu H|u_n\rangle\langle u_{m}^{L}|\partial_\nu H|u_n\rangle|}{\left(E^L_{m}-E_n\right)\left(E_{m}-E_n^L\right)}$, respectively, where $|u_{m,n}\rangle$ and $\langle u_{m,n}^L|$ denote the right and left eigenstates.
\begin{figure}[t!] 
		\centering
		\includegraphics[width=3.3in]{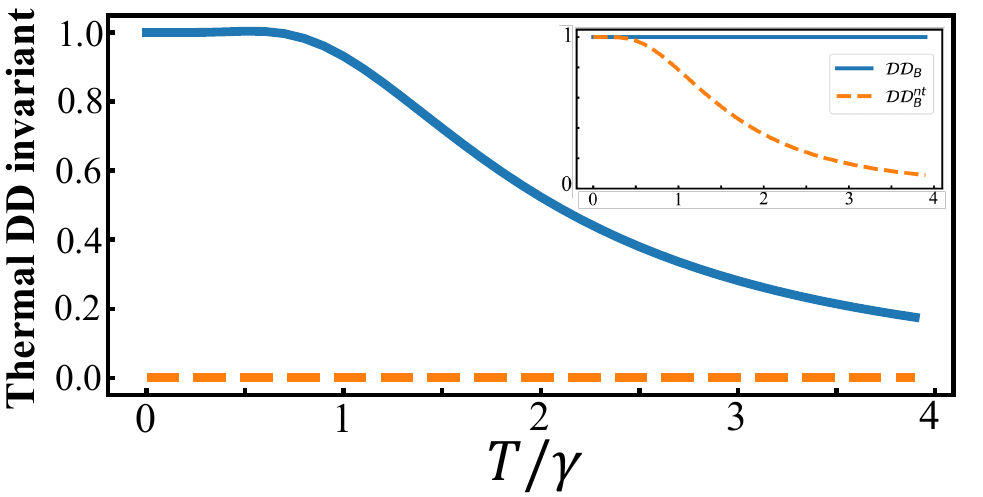}
		\caption{The NT thermal DD invariant $\mathcal{DD}_B^{nt}$ versus temperature $T$ (scales with $\gamma$). The blue solid line represents the 4D parameter space enclosing the ES, while the yellow dash line corresponds to the case while the parameter space without enclosing the ES. The inset shows the Hermitian case, where the blue solid and yellow dash lines are the $\mathcal{DD}_B$ and $\mathcal{DD}_B^{nt}$ as functions of temperature $T$, respectively.}
		\label{Fig_DDNH}
\end{figure}%
    For mixed states, we propose the thermal DD invariant, which can be obtained from the Bures metric \cite{HUBNER1993226}
    \begin{eqnarray}
        \mathcal{G}_B^{\mu\nu}=\frac{1}{2}\sum_{m,n}\frac{\langle u_m|\partial_\mu\rho|u_n\rangle\langle u_n|\partial_\nu \rho|u_m\rangle}{P_m+P_n}, 
    \end{eqnarray}
    where $P_{m}$ and $P_{n}$ are the coefficients of the system's initial state. In the Hermitian case,
    the mixed-state density matrix is given as
    \begin{eqnarray}
        \rho = P_0|u_0\rangle\langle u_0|+P_+|u_+\rangle\langle u_+|+P_-|u_-\rangle\langle u_-|,
    \end{eqnarray}
    where $P_0=1/Z$ and $P_\pm=e^{\mp E_+/T}/Z$. The thermal three-form Berry curvature, calculated via the Bures metric, reads
    \begin{eqnarray}
        \mathcal{M}_B&&=4\sqrt{\mathcal{G}_B^{\alpha\alpha}\left(\mathcal{G}_B^{\phi_1\phi_1}\mathcal{G}_B^{\phi_2\phi_2}-|\mathcal{G}_B^{\phi_1\phi_2}|^2\right)},
    \end{eqnarray}
    from which we introduce the thermal DD invariant
    \begin{eqnarray}
        \mathcal{DD}_B=\frac{1}{2\pi^2}\int_{S^3}\lambda_1\left(E,T\right)\mathcal{M}_B,
    \end{eqnarray}
    with $\lambda_1\left(E,T\right)=\frac{2\sqrt{2}\sin{2\alpha}\sinh^2{\left(E_+/2T\right)}\sinh{\left(E_+/T\right)}}{\sqrt{\cosh{\left(E_+/T\right)}\left(1+2\cosh{\left(E_+/T\right)}\right)^3}}$, yielding $\mathcal{DD}_B=1$. Analogously, we introduce the NT thermal DD invariant as
    \begin{eqnarray}
        \mathcal{DD}_B^{nt}=\frac{1}{2\pi^2}\int_{S^3}\mathcal{M}_B,
    \end{eqnarray}
    whose temperature dependence is shown in the inset of Fig.~\ref{Fig_DDNH}. The blue solid line represents $\mathcal{DD}_B$, which remain unity at finite temperature, while the yellow dash line corresponds to $\mathcal{DD}_B^{nt}$, which gradually decreases as the temperature rises.

    In the NH case,
    the mixed-state density matrix at temperature $T$ is described as $\rho=\sum_{n}^{1,2,3}P_n|u_n\rangle\langle u_n^L|$, where $P_n=e^{- E_n/T}/Z$, and the Bures metric is rewritten as
    \begin{eqnarray}
        \mathcal{G}_B^{\prime\mu\nu}&=&\sum_{m,n}\left(P_m\langle\partial_\mu u_m|u_n\rangle+P_n\langle u_m|\partial_\mu u_n\rangle\right)\nonumber\\
        &&\times\left(P_m\langle u_n|\partial_\nu u_m\rangle+P_n\langle\partial_\nu u_n| u_m\rangle\right)/\left(P_m+P_n\right).\nonumber\\
    \end{eqnarray}
    As the temperature $T$ varies, the calculated NT thermal DD invariant $\mathcal{DD}_B^{nt}$ versus $T$ is plotted against $T$ in Fig.~\ref{Fig_DDNH}.
    The blue solid curve shows the result when the parameter sphere encloses the ES, in which case $\mathcal{DD}_B^{nt}$ decreases from unity as $T$ increases. The yellow dashed line shows $\mathcal{DD}_B^{nt}$ versus $T$ when the parameter sphere does not enclose the ES, where it remains zero across the entire temperature range. A topological transition, characterized by the thermal DD invariant, occurs when the parameter space crosses the ES.  
    
	\section{Mixed-State Topology of the 4D non-Abelian system}

    We next investigate the mixed-state topology of the 4D non-Abelian system using the Uhlmann phase and the second Chern number \cite{sugawa_second_2018,yang2025hyperspherelikenonabelianyangmonopole}. 
    
	\subsection{The Uhlmann phase}
	We consider a 4D non-Abelian system with particle gain and loss, described by the Hamiltonian
	\begin{eqnarray}
		H_{4}&=&\sum_{\mu=1}^{5}q_\mu\Gamma_\mu+i\gamma\Gamma_4,
        \label{eq24}
	\end{eqnarray}
	where $\Gamma_\mu$ are fourth-order Dirac matrices satisfying the Clifford algebra $\{\Gamma_m,\Gamma_n\}=2\delta_{mn}I_0^{4*4}$ \cite{yang2025hyperspherelikenonabelianyangmonopole},
	and $\gamma$ quantifies the strength of gain and loss. When $q_4=0$, the Hamiltonian (\ref{eq24}) hosts two degenerate eigenenergies $E_{\pm}=\sqrt{|q|^2-\gamma^2}$, which coalesce at $q_1^2+q_2^2+q_3^2+q_5^2=\gamma^2$, an exceptional hypersphere (EHS) on the $\{q_1,q_2,q_3,q_5\}$ subspace. 
	
	To probe the Uhlmann phase associated with such a EHS, we construct a parameterized evolution path for $\vec{q}=\{(r\sin{\theta}+d)/\sqrt{2},(r\sin{\theta}+d)/\sqrt{2},0,r\cos{\theta},0\}$, with $r/\gamma=2$ and $d/\gamma=5/2$. This parameter loop lies in the $\{q_1,q_3,q_5\}$ subspace and intertwines with the EHS. The eigenenergies are given by $E_{1,2}=\pm\sqrt{r^2+d^2-\gamma^2+2rd\sin{\theta}+2ir\gamma\cos{\theta}}$,
	and the degenerate right eigenstates are $|u_{1,2}^\alpha\rangle=\left[
	 		0,r\sin{\theta}+d,
	 		\sqrt{2}\left(E_{1,2}+r\cos{\theta}+i\gamma\right),r\sin{\theta}+d
	 	\right]^T/N_{1,2}$ and $|u_{1,2}^\beta\rangle=
	 	\left[
	 		\sqrt{2}\left(E_{1,2}+r\cos{\theta}+i\gamma\right),
	 		r\sin{\theta}+d,
	 		0,\notag \right.
\\
\left.-(r\sin{\theta}+d)
	 	\right]^T/N_{1,2}$.
\begin{figure}[t!] 
		\centering
		\includegraphics[width=3.3in]{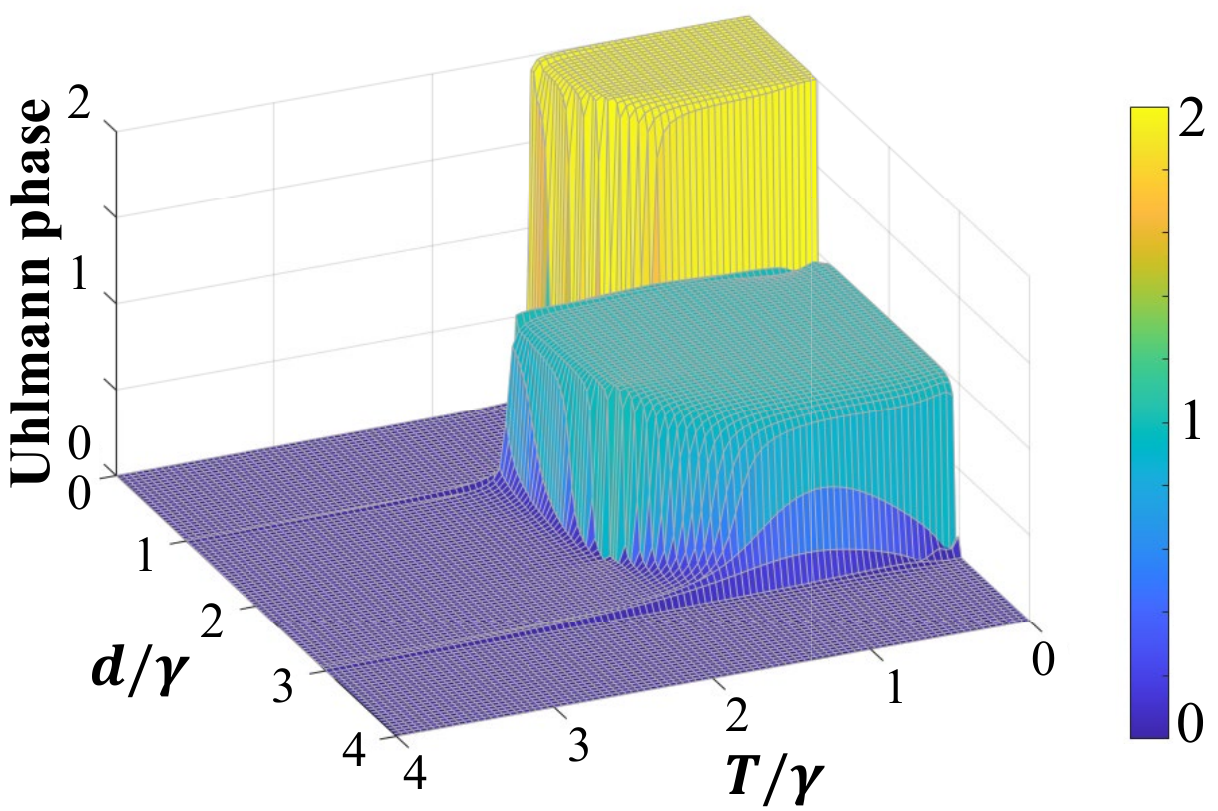}
		\caption{The Uhlmann phase $\Phi_U$ as functions of temperature $T$ and parameter loop displacement $d$, both scale with $\gamma$.}
		\label{Fig5}
	\end{figure}%
	The corresponding degenerate left eigenstates $\langle u^{L\alpha\beta}_{1,2}|$ are obtained by the biorthogonal condition $\langle u_{j,k}^{L{m,n}}|u_{j,k}^{m,n}\rangle=\delta_{jk}\delta_{mn}$. At finite temperature $T$, the mixed-state density matrix is expressed as
	\begin{eqnarray}
		\rho=\sum_{n=1,2}P_n\left(|u_n^\alpha\rangle\langle u_n^{L\alpha}|+|u_n^\beta\rangle\langle u_n^{L\beta}|\right),
		\label{eq17}
	\end{eqnarray}
	where $P_n$ represents the Boltzmann weights. The Uhlmann connection is given by
	\begin{small}
	\begin{equation}
    A_U^\theta=\sum_{m,n}^{1,2}\sum_{j,k}^{\alpha,\beta}\frac{|u_m^j(\theta)\rangle\langle u^{Lj}_m(\theta)|\left[\partial_\theta\sqrt{\rho_\theta},\sqrt{\rho_\theta}\right]|u_n^k(\theta)\rangle\langle u^{Lk}_n(\theta)|}{P_m(\theta)+P_n(\theta)}\textrm{d}\theta.\\
		\label{eq18}
	\end{equation}
	\end{small}%
	Applying the parallel transport condition to Eqs.~(\ref{eq17}) and (\ref{eq18}) under the biorthogonal constraints yields
	\begin{eqnarray}
		A_U^\theta&=&\sum_{m\neq n}^{1,2}\sum_{j,k}^{\alpha,\beta}\left[f(T)-f^2(T)\right]\langle\partial_\theta u_m^{Lj}|u_n^k\rangle|u_m^{Lj}\rangle\langle u_n^k|\nonumber\\
		&&+\sum_{m\neq n}^{1,2}\sum_{j,k}^{\alpha,\beta}\left[3f(T)-f^2(T)\right]\langle u_m^{Lj}|\partial_\theta u_n^k\rangle|u_m^{Lj}\rangle\langle u_n^k|.\nonumber\\
	\end{eqnarray}
	Crucially, the mixed state in Eq.~(\ref{eq17}) must complete two winding loops $2\mathcal{L}$ to accumulate a well-defined Uhlmann phase
	\begin{eqnarray}
		\Phi_U=\textrm{arg Tr}(\rho_0 e^{\int A_U^\theta\textrm{d}\theta})=\pi.
	\end{eqnarray}
	As shown in Fig.~\ref{Fig5}, the Uhlmann phase $\Phi_U$ drops abruptly to $0$ when the temperature 
	exceeds the critical value of $T/\gamma=1.98$. Meanwhile, for a fixed $T$ as $d/\gamma$ increases from 0 to 4, the Uhlmann phase exhibts twofold distinct topological transitions.
	
	\subsection{The second thermal Uhlmann-Chern number}
	To characterize the mixed-state topology via the second thermal Uhlmann-Chern number, we consider the 4D NH Hamiltonian in Eq.~(\ref{eq24}), which is associated with the five-dimensional (5D) hypersphere in the parameter space spanned by $\{\theta_1,\theta_2,\phi_1,\phi_2$, $R$\}, where $R$ is the radius of the hypersphere. The generic form of $\vec{q}$ is taken as 
    $\vec{q}=R\{\sin\theta_1\sin\theta_2\cos\phi_2,\sin\theta_1\cos\theta_2\cos\phi_1,\sin\theta_1\cos\theta_2\sin\phi_1,\\
    \cos\theta_1,\sin\theta_1\sin\theta_2\sin\phi_2\}$ in Eq.~(\ref{eq24}).
	The eigenenergies and biorthogonal eigenstates are given by $E_{1,2}=\pm\sqrt{R^2-\gamma^2+2i\gamma R\cos{\theta_1}}$,
    and $|u_{1,2}^\alpha\rangle=
		\left[
			0,
			R\sin{\theta_1}\sin{\theta_2}e^{-i(\phi_1-\phi_2)},
			(E_{1,2}+R\cos{\theta}+i\gamma)e^{-i\phi_1},\notag \right.
\\
\left.
			R\sin{\theta_1}\cos{\theta_2}
		\right]/N_{1,2}$, $|u_{1,2}^\beta\rangle=
		\left[
			(E_{1,2}+R\cos{\theta}+i\gamma)e^{i\phi_2},\right.$\\
            $\left. R\sin{\theta_1}\cos{\theta_2}e^{-i(\phi_1-\phi_2)},
			0,
			-R\sin{\theta_1}\sin{\theta_2}
		\right]/N_{1,2}$,
	with the corresponding degenerate left eigenstates $\langle u^{L\alpha\beta}_{1,2}|$. 
	
	At finite temperature $T$, the four components of Uhlmann connection derived from Eqs.~(\ref{eq17}) and (\ref{eq18}), which extend the parallel transport condition to mixed states under non-Abelian gauge symmetry, can be expressed as
	\begin{eqnarray}
		A_U^{\theta_1}&=&f(T)\sum_{j=\alpha,\beta}\langle u_1^j|\partial_{\theta_1}u_2^j\rangle\left(|u_1^j\rangle\langle u_2^j|-|u_2^j\rangle\langle u_1^j|\right),\nonumber\\
		A_U^{\theta_2}&=&-f(T)\frac{R^2\sin^2{\theta_1}}{N_1N_2}\sum_{m\ne n}^{1,2}\left(|u_m^\alpha\rangle\langle u_n^\beta|-|u_n^\beta\rangle\langle u_m^\alpha|\right),\nonumber\\
		A_U^{\phi_1}&=&-f(T)\frac{R^2\sin^2{\theta_1}}{2N_1N_2}\sum_{m\ne n}^{1,2}\left[\sum_{j\ne k}^{\alpha,\beta}\sin{2\theta_2}|u_m^j\rangle\langle u_n^k|\right.\nonumber\\
		&&\left. -2\cos^2{\theta_2}\left(|u_m^\alpha\rangle\langle u_m^\alpha|+|u_m^\beta\rangle\langle u_m^\beta|\right)\right],\nonumber\\
		A_U^{\phi_2}&=&f(T)\frac{R^2\sin^2{\theta_1}}{2N_1N_2}\sum_{m\ne n}^{1,2}\left[\sum_{j\ne k}^{\alpha,\beta}\sin{2\theta_2}|u_m^j\rangle\langle u_n^k|\right.\nonumber\\
		&&\left. +2\sin^2{\theta_2}\left(|u_m^\alpha\rangle\langle u_m^\alpha|-|u_m^\beta\rangle\langle u_m^\beta|\right)\right].
	\end{eqnarray}
    \begin{figure}[t!] 
		\centering
		\includegraphics[width=3.3in]{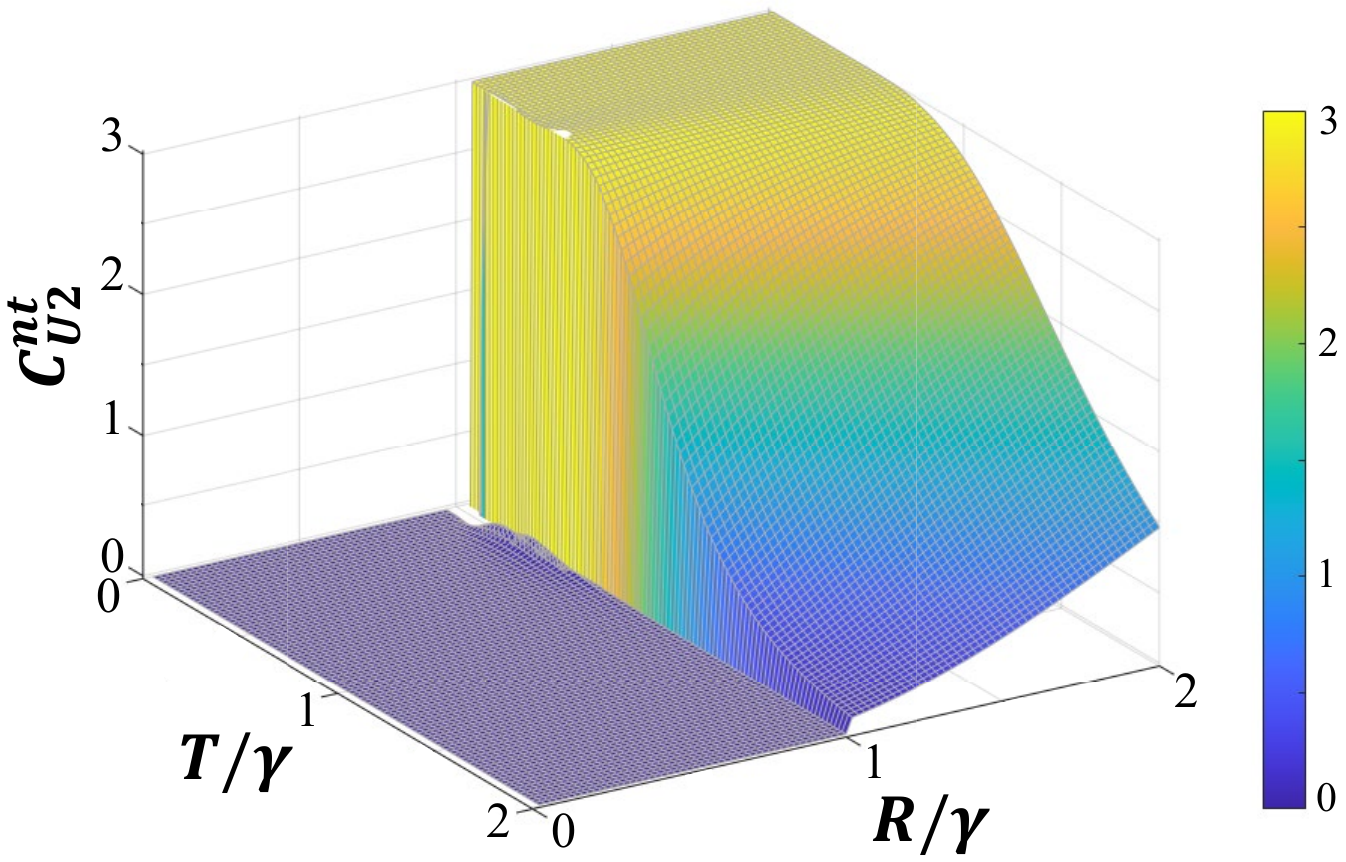}
		\caption{The NT thermal Uhlmann-Chern number $C_{U2}^{nt}$ versus temperature $T$ and parameter hypersphere radius R, both scale with $\gamma$.}
		\label{Fig6}
	\end{figure}%
	Substituting the density matrix from Eq.~(\ref{eq17}) into the Uhlmann curvature yields
	\begin{eqnarray}
		\textrm{Tr}(\rho F_U^{\theta_1\theta_2})=\textrm{Tr}(\rho F_U^{\phi_1\phi_2})=0,
	\end{eqnarray}
	and calculating the first thermal Uhlmann-Chern number gives
	\begin{eqnarray}
		C_{U1} = \frac{i}{2\pi}\int \lambda(E,T)\textrm{Tr}(\rho F_U)=C_{U1}^{nt}=0,
		\label{eq27}
	\end{eqnarray}
	which is consistent with results for zero-temperature non-Abelian Hermitian systems. 
	
	The second thermal Uhlmann-Chern number $C_{U2}$ reveals a novel feature. Its general expression is
	\begin{align}
		C_{U2} = \frac{1}{8\pi^2}\int \lambda_2(E,T)\textrm{Tr}(\rho F_U\wedge F_U),
		\label{eq28}
	\end{align}
	where $F_U$ is the Uhlmann curvature tensor and $1/\lambda_2(E,T)=\tanh^5{\frac{E}{T}}$. To compute $C_{U2}$, we explicitly evaluate all components of the Uhlmann curvature as follows:
	\begin{widetext}
	\begin{eqnarray}
		&&\textrm{Tr}\left[\rho\left(\partial_{\phi_1}A_U^{\phi_2}-\partial_{\phi_2}A_U^{\phi_1}\right)\left(\partial_{\theta_1}A_U^{\theta_2}-\partial_{\theta_2}A_U^{\theta_1}\right)\right]=4\tanh{\frac{E}{T}}f^2(T)\frac{R^6\sin{\theta_1}^6\sin{2\theta_2}}{(N_1N_2)^3}\left(\langle u_1^\alpha|\partial_{\theta_1}u_2^\alpha\rangle+\langle u_1^\beta|\partial_{\theta_1}u_2^\beta\rangle\right),\nonumber\\
		&&\textrm{Tr}\left[\rho\left(\partial_{\phi_1}A_U^{\phi_2}-\partial_{\phi_2}A_U^{\phi_1}\right)\left(A_U^{\theta_1}A_U^{\theta_2}-A_U^{\theta_2}A_U^{\theta_1}\right)\right]=-2\tanh{\frac{E}{T}}f^3(T)\frac{R^6\sin{\theta_1}^6\sin{2\theta_2}}{(N_1N_2)^3}\left(\langle u_1^\alpha|\partial_{\theta_1}u_2^\alpha\rangle+\langle u_1^\beta|\partial_{\theta_1}u_2^\beta\rangle\right),\nonumber\\
		&&\textrm{Tr}\left[\rho\left(A_U^{\phi_1}A_U^{\phi_2}-A_U^{\phi_2}A_U^{\phi_1}\right)\left(\partial_{\theta_1}A_U^{\theta_2}-\partial_{\theta_2}A_U^{\theta_1}\right)\right]=-2\tanh{\frac{E}{T}}f^3(T)\frac{R^6\sin{\theta_1}^6\sin{2\theta_2}}{(N_1N_2)^3}\left(\langle u_1^\alpha|\partial_{\theta_1}u_2^\alpha\rangle+\langle u_1^\beta|\partial_{\theta_1}u_2^\beta\rangle\right),\nonumber\\
		&&\textrm{Tr}\left[\rho\left(A_U^{\phi_1}A_U^{\phi_2}-A_U^{\phi_2}A_U^{\phi_1}\right)\left(A_U^{\theta_1}A_U^{\theta_2}-A_U^{\theta_2}A_U^{\theta_1}\right)\right]=\tanh{\frac{E}{T}}f^4(T)\frac{R^6\sin{\theta_1}^6\sin{2\theta_2}}{(N_1N_2)^3}\left(\langle u_1^\alpha|\partial_{\theta_1}u_2^\alpha\rangle+\langle u_1^\beta|\partial_{\theta_1}u_2^\beta\rangle\right).\nonumber\\
	\end{eqnarray}
	\end{widetext}
	By symmetry, $\textrm{Tr}(\rho F_U^{\phi_1\phi_2} F_U^{\theta_1\theta_2}-\rho F_U^{\phi_1\theta_1} F_U^{\phi_2\theta_2}+\rho F_U^{\phi_1\theta_2} F_U^{\phi_2\theta_1})=3\textrm{Tr}(\rho F_U^{\phi_1\phi_2} F_U^{\theta_1\theta_2})$, we finally obtain $C_{U2}=3$ when $R/\gamma>1$ and $C_{U2}=0$ when $R/\gamma<1$. Furthermore, the NT (non-topological) second Uhlmann-Chern number $C_{U2}^{nt}$ is given by
	\begin{eqnarray}
		C_{U2}^{nt}=24R^6\tanh^5{\frac{E}{T}}\int \frac{\sin^6{\theta_1}}{(N_1N_2)^3}\langle u_1^\alpha|\partial_{\theta_1}u_2^\alpha\rangle\textrm{d}\theta_1. 
	\end{eqnarray}
	In Fig.~\ref{Fig6}, the NT thermal Uhlmann-Chern number $C_{U2}^{nt}$ is plotted with respect to the the parameter hypersphere radius $R$ and the temperature $T$, both scaled by $\gamma$. The result shows that $C_{U2}^{nt}$ decreases gradually with increasing thermal noise, but abruptly drops to zero when the radius $R$ is reduced to the regime $R/\gamma<1$, where the parameter hypersphere no longer encloses the EHS.

    \section{EXPERIMENTAL feasibility}
    Characterizing the mixed-state topology requires three steps. First, the eigenenergies and eigenstates of the NH systems can be extracted using the method from our previous 2D/3D implementations \cite{PhysRevLett.131.260201,han_measuring_2024} or the protocols proposed for higher dimensions \cite{s11433_025_2851_8,yang2025hyperspherelikenonabelianyangmonopole}. Second, at a specific (unknown) temperature, the system is evolved to the thermal equilibrium, and the thermal state is reconstructed via quantum state tomography. The Boltzmann weights are then derived from the tomographic data combined with the extracted eigenenergies and eigenstates. Third, all the thermal topological invariants are computed from the the obtained quantities.

	\section{conclusion}
    We have investigated the mixed-state topology in 2D, 3D and 4D NH systems, associated with the ER, ES and EHS, respectively, which are characterized by the thermal Uhlmann-Chern number, the thermal DD invariant and the second thermal Uhlmann-Chern number. Our results reveal distinctive mixed-state topology absent in pure-state cases, thereby significantly extending the scope of NH topology.     

    This work was supported by the National Natural Science Foundation of China (Grant Nos. 12475015, 12474356, 12274080, 1187510).

\bibliography{NHmixedstate}

@article{RevModPhys.82.3045,
  title = {Colloquium: Topological insulators},
  author = {Hasan, M. Z. and Kane, C. L.},
  journal = {Rev. Mod. Phys.},
  volume = {82},
  issue = {4},
  pages = {3045--3067},
  numpages = {0},
  year = {2010},
  month = {Nov},
  publisher = {American Physical Society},
  doi = {10.1103/RevModPhys.82.3045},
  url = {https://link.aps.org/doi/10.1103/RevModPhys.82.3045}
}

@article{RevModPhys.83.1057,
  title = {Topological insulators and superconductors},
  author = {Qi, Xiao-Liang and Zhang, Shou-Cheng},
  journal = {Rev. Mod. Phys.},
  volume = {83},
  issue = {4},
  pages = {1057--1110},
  numpages = {0},
  year = {2011},
  month = {Oct},
  publisher = {American Physical Society},
  doi = {10.1103/RevModPhys.83.1057},
  url = {https://link.aps.org/doi/10.1103/RevModPhys.83.1057}
}

@book{shentopological2012,
	address = {Berlin, Heidelberg},
	series = {Springer Series in Solid-State Sciences},
	title = {Topological Insulators: Dirac Equation in Condensed Matters},
	volume = {174},
	urldate = {2025-02-19},
	publisher = {Springer Berlin Heidelberg},
	author = {Shen, Shun-Qing},
	year = {2012},
	doi = {10.1007/978-3-642-32858-9}
}

@book{bernevigtopological2013,
	address = {Princeton},
	title = {Topological Insulators and Topological Superconductors},
	publisher = {Princeton University Press},
	author = {Bernevig, B. Andrei},
	collaborator = {Hughes, Taylor L.},
	year = {2013}
}

@article{PhysRevLett.61.2015,
  title = {Model for a Quantum Hall Effect without Landau Levels: Condensed-Matter Realization of the "Parity Anomaly"},
  author = {Haldane, F. D. M.},
  journal = {Phys. Rev. Lett.},
  volume = {61},
  issue = {18},
  pages = {2015--2018},
  numpages = {0},
  year = {1988},
  month = {Oct},
  publisher = {American Physical Society},
  doi = {10.1103/PhysRevLett.61.2015},
  url = {https://link.aps.org/doi/10.1103/PhysRevLett.61.2015}
}

@article{RevModPhys.88.035005,
  title = {Classification of topological quantum matter with symmetries},
  author = {Chiu, Ching-Kai and Teo, Jeffrey C. Y. and Schnyder, Andreas P. and Ryu, Shinsei},
  journal = {Rev. Mod. Phys.},
  volume = {88},
  issue = {3},
  pages = {035005},
  numpages = {63},
  year = {2016},
  month = {Aug},
  publisher = {American Physical Society},
  doi = {10.1103/RevModPhys.88.035005},
  url = {https://link.aps.org/doi/10.1103/RevModPhys.88.035005}
}

@article{PhysRevLett.95.146802,
  title = {Z2 Topological Order and the Quantum Spin Hall Effect},
  author = {Kane, C. L. and Mele, E. J.},
  journal = {Phys. Rev. Lett.},
  volume = {95},
  issue = {14},
  pages = {146802},
  numpages = {4},
  year = {2005},
  month = {Sep},
  publisher = {American Physical Society},
  doi = {10.1103/PhysRevLett.95.146802},
  url = {https://link.aps.org/doi/10.1103/PhysRevLett.95.146802}
}

@article{science.1133734,
author = {B. Andrei Bernevig  and Taylor L. Hughes  and Shou-Cheng Zhang },
title = {Quantum Spin Hall Effect and Topological Phase Transition in HgTe Quantum Wells},
journal = {Science},
volume = {314},
number = {5806},
pages = {1757-1761},
year = {2006},
doi = {10.1126/science.1133734},
url = {https://www.science.org/doi/abs/10.1126/science.1133734}
}

@article{science.1148047,
author = {Markus König  and Steffen Wiedmann  and Christoph Brüne  and Andreas Roth  and Hartmut Buhmann  and Laurens W. Molenkamp  and Xiao-Liang Qi  and Shou-Cheng Zhang },
title = {Quantum Spin Hall Insulator State in HgTe Quantum Wells},
journal = {Science},
volume = {318},
number = {5851},
pages = {766-770},
year = {2007},
doi = {10.1126/science.1148047},
url = {https://www.science.org/doi/abs/10.1126/science.1148047}
}

@article{PhysRevLett.98.106803,
  title = {Topological Insulators in Three Dimensions},
  author = {Fu, Liang and Kane, C. L. and Mele, E. J.},
  journal = {Phys. Rev. Lett.},
  volume = {98},
  issue = {10},
  pages = {106803},
  numpages = {4},
  year = {2007},
  month = {Mar},
  publisher = {American Physical Society},
  doi = {10.1103/PhysRevLett.98.106803},
  url = {https://link.aps.org/doi/10.1103/PhysRevLett.98.106803}
}

@article{PhysRevB.76.045302,
  title = {Topological insulators with inversion symmetry},
  author = {Fu, Liang and Kane, C. L.},
  journal = {Phys. Rev. B},
  volume = {76},
  issue = {4},
  pages = {045302},
  numpages = {17},
  year = {2007},
  month = {Jul},
  publisher = {American Physical Society},
  doi = {10.1103/PhysRevB.76.045302},
  url = {https://link.aps.org/doi/10.1103/PhysRevB.76.045302}
}

@article{hsiehtopological2008,
	title = {A topological {Dirac} insulator in a quantum spin {Hall} phase},
	volume = {452},
	issn = {0028-0836, 1476-4687},
	url = {https://www.nature.com/articles/nature06843},
	doi = {10.1038/nature06843},
	language = {en},
	number = {7190},
	urldate = {2023-08-08},
	journal = {Nature},
	author = {Hsieh, D. and Qian, D. and Wray, L. and Xia, Y. and Hor, Y. S. and Cava, R. J. and Hasan, M. Z.},
	month = apr,
	year = {2008},
	pages = {970--974}
}

@article{noauthorquantal1984,
	title = {Quantal phase factors accompanying adiabatic changes},
	volume = {392},
	issn = {0080-4630},
	url = {https://royalsocietypublishing.org/doi/10.1098/rspa.1984.0023},
	doi = {10.1098/rspa.1984.0023},
	number = {1802},
	urldate = {2025-02-20},
	journal = {Proc. R. Soc. A},
	author = {Berry, M. V.},
	month = mar,
	year = {1984},
	pages = {45--57}
}

@article{rayobservation2014,
	title = {Observation of Dirac monopoles in a synthetic magnetic field},
	volume = {505},
	issn = {0028-0836, 1476-4687},
	url = {https://www.nature.com/articles/nature12954},
	doi = {10.1038/nature12954},
	number = {7485},
	urldate = {2025-02-20},
	journal = {Nature},
	author = {Ray, M. W. and Ruokokoski, E. and Kandel, S. and Möttönen, M. and Hall, D. S.},
	month = jan,
	year = {2014},
	pages = {657--660}
}

@article{noauthorquantised1931,
	title = {Quantised singularities in the electromagnetic field},
	volume = {133},
	issn = {0950-1207, 2053-9150},
	url = {https://royalsocietypublishing.org/doi/10.1098/rspa.1931.0130},
	doi = {10.1098/rspa.1931.0130},
	number = {821},
	urldate = {2025-02-20},
	journal = {Proc. R. Soc. Lond. Ser. A},
	author = {P. A. M. Dirac},
	month = sep,
	year = {1931},
	pages = {60--72}
}

@article{sawadasemi-empirical1974,
	title = {Semi-empirical detection of the Dirac's magnetic monopoles},
	volume = {52},
	issn = {03702693},
	url = {https://linkinghub.elsevier.com/retrieve/pii/0370269374907217},
	doi = {10.1016/0370-2693(74)90721-7},
	number = {1},
	urldate = {2025-02-20},
	journal = {Phys. Lett. B},
	author = {Sawada, T.},
	month = sep,
	year = {1974},
	pages = {67--70}
}

@article{jackiwdiracs2004,
	title = {Dirac's Magnetic Monopoles (Again)},
	volume = {19},
	issn = {0217-751X, 1793-656X},
	url = {https://www.worldscientific.com/doi/abs/10.1142/S0217751X04018658},
	doi = {10.1142/S0217751X04018658},
	number = {supp01},
	journal = {Int. J. Mod. Phys. A},
	author = {Jackiw, Roman W.},
	month = feb,
	year = {2004},
	pages = {137--143}
}

@article{PhysRevLett.113.050402,
  title = {Measuring a Topological Transition in an Artificial Spin-$1/2$ System},
  author = {Schroer, M. D. and Kolodrubetz, M. H. and Kindel, W. F. and Sandberg, M. and Gao, J. and Vissers, M. R. and Pappas, D. P. and Polkovnikov, Anatoli and Lehnert, K. W.},
  journal = {Phys. Rev. Lett.},
  volume = {113},
  issue = {5},
  pages = {050402},
  numpages = {5},
  year = {2014},
  month = {Jul},
  publisher = {American Physical Society},
  doi = {10.1103/PhysRevLett.113.050402},
  url = {https://link.aps.org/doi/10.1103/PhysRevLett.113.050402}
}

@article{roushan_observation_2014,
	title = {Observation of topological transitions in interacting quantum circuits},
	volume = {515},
	issn = {0028-0836, 1476-4687},
	url = {https://www.nature.com/articles/nature13891},
	doi = {10.1038/nature13891},
	number = {7526},
	urldate = {2025-02-20},
	journal = {Nature},
	author = {Roushan, P. and Neill, C. and Chen, Yu and Kolodrubetz, M. and Quintana, C. and Leung, N. and Fang, M. and Barends, R. and Campbell, B. and Chen, Z. and Chiaro, B. and Dunsworth, A. and Jeffrey, E. and Kelly, J. and Megrant, A. and Mutus, J. and O’Malley, P. J. J. and Sank, D. and Vainsencher, A. and Wenner, J. and White, T. and Polkovnikov, A. and Cleland, A. N. and Martinis, J. M.},
	month = nov,
	year = {2014},
	pages = {241--244},
}

@article{PhysRevLett.126.017702,
  title = {Experimental Observation of Tensor Monopoles with a Superconducting Qudit},
  author = {Tan, Xinsheng and Zhang, Dan-Wei and Zheng, Wen and Yang, Xiaopei and Song, Shuqing and Han, Zhikun and Dong, Yuqian and Wang, Zhimin and Lan, Dong and Yan, Hui and Zhu, Shi-Liang and Yu, Yang},
  journal = {Phys. Rev. Lett.},
  volume = {126},
  issue = {1},
  pages = {017702},
  numpages = {6},
  year = {2021},
  month = {Jan},
  publisher = {American Physical Society},
  doi = {10.1103/PhysRevLett.126.017702},
  url = {https://link.aps.org/doi/10.1103/PhysRevLett.126.017702}
}

@article{chen_synthetic_2022,
	title = {A synthetic monopole source of Kalb-Ramond field in diamond},
	volume = {375},
	issn = {0036-8075, 1095-9203},
	url = {https://www.science.org/doi/10.1126/science.abe6437},
	doi = {10.1126/science.abe6437},
	number = {6584},
	urldate = {2025-02-20},
	journal = {Science},
	author = {Chen, Mo and Li, Changhao and Palumbo, Giandomenico and Zhu, Yan-Qing and Goldman, Nathan and Cappellaro, Paola},
	month = mar,
	year = {2022},
	pages = {1017--1020}
}

@article{sugawa_second_2018,
	title = {Second {Chern} number of a quantum-simulated non-{Abelian} {Yang} monopole},
	volume = {360},
	issn = {0036-8075, 1095-9203},
	url = {https://www.science.org/doi/10.1126/science.aam9031},
	doi = {10.1126/science.aam9031},
	number = {6396},
	urldate = {2025-02-20},
	journal = {Science},
	author = {Sugawa, Seiji and Salces-Carcoba, Francisco and Perry, Abigail R. and Yue, Yuchen and Spielman, I. B.},
	month = jun,
	year = {2018},
	pages = {1429--1434}
}

@article{PhysRevA.98.013603,
  title = {Topological phase transitions in tilted optical lattices},
  author = {Kolovsky, Andrey R.},
  journal = {Phys. Rev. A},
  volume = {98},
  issue = {1},
  pages = {013603},
  numpages = {4},
  year = {2018},
  month = {Jul},
  publisher = {American Physical Society},
  doi = {10.1103/PhysRevA.98.013603},
  url = {https://link.aps.org/doi/10.1103/PhysRevA.98.013603}
}

@article{PhysRevA.89.013627,
  title = {Nonequilibrium topological phase transitions in two-dimensional optical lattices},
  author = {Nakagawa, Masaya and Kawakami, Norio},
  journal = {Phys. Rev. A},
  volume = {89},
  issue = {1},
  pages = {013627},
  numpages = {8},
  year = {2014},
  month = {Jan},
  publisher = {American Physical Society},
  doi = {10.1103/PhysRevA.89.013627},
  url = {https://link.aps.org/doi/10.1103/PhysRevA.89.013627}
}

@article{PhysRevLett.125.217202,
  title = {Bosonic Bott Index and Disorder-Induced Topological Transitions of Magnons},
  author = {Wang, X. S. and Brataas, Arne and Troncoso, Roberto E.},
  journal = {Phys. Rev. Lett.},
  volume = {125},
  issue = {21},
  pages = {217202},
  numpages = {6},
  year = {2020},
  month = {Nov},
  publisher = {American Physical Society},
  doi = {10.1103/PhysRevLett.125.217202},
  url = {https://link.aps.org/doi/10.1103/PhysRevLett.125.217202}
}

@article{PhysRevA.98.033830,
  title = {Topological phase transition in a stretchable photonic crystal},
  author = {Saei Ghareh Naz, Ehsan and Fulga, Ion Cosma and Ma, Libo and Schmidt, Oliver G. and van den Brink, Jeroen},
  journal = {Phys. Rev. A},
  volume = {98},
  issue = {3},
  pages = {033830},
  numpages = {9},
  year = {2018},
  month = {Sep},
  publisher = {American Physical Society},
  doi = {10.1103/PhysRevA.98.033830},
  url = {https://link.aps.org/doi/10.1103/PhysRevA.98.033830}
}

@article{PhysRevA.97.031801,
  title = {Photonic realization of a transition to a strongly driven Floquet topological phase},
  author = {Guglielmon, Jonathan and Huang, Sheng and Chen, Kevin P. and Rechtsman, Mikael C.},
  journal = {Phys. Rev. A},
  volume = {97},
  issue = {3},
  pages = {031801},
  numpages = {5},
  year = {2018},
  month = {Mar},
  publisher = {American Physical Society},
  doi = {10.1103/PhysRevA.97.031801},
  url = {https://link.aps.org/doi/10.1103/PhysRevA.97.031801}
}

@article{PhysRevLett.119.183901,
  title = {Disorder-Induced Topological State Transition in Photonic Metamaterials},
  author = {Liu, Changxu and Gao, Wenlong and Yang, Biao and Zhang, Shuang},
  journal = {Phys. Rev. Lett.},
  volume = {119},
  issue = {18},
  pages = {183901},
  numpages = {5},
  year = {2017},
  month = {Nov},
  publisher = {American Physical Society},
  doi = {10.1103/PhysRevLett.119.183901},
  url = {https://link.aps.org/doi/10.1103/PhysRevLett.119.183901}
}

@article{PhysRevLett.86.787,
  title = {Experimental Observation of the Topological Structure of Exceptional Points},
  author = {Dembowski, C. and Gr\"af, H.-D. and Harney, H. L. and Heine, A. and Heiss, W. D. and Rehfeld, H. and Richter, A.},
  journal = {Phys. Rev. Lett.},
  volume = {86},
  issue = {5},
  pages = {787--790},
  numpages = {0},
  year = {2001},
  month = {Jan},
  publisher = {American Physical Society},
  doi = {10.1103/PhysRevLett.86.787},
  url = {https://link.aps.org/doi/10.1103/PhysRevLett.86.787}
}

@article{PhysRevLett.104.153601,
  title = {Quasieigenstate Coalescence in an Atom-Cavity Quantum Composite},
  author = {Choi, Youngwoon and Kang, Sungsam and Lim, Sooin and Kim, Wookrae and Kim, Jung-Ryul and Lee, Jai-Hyung and An, Kyungwon},
  journal = {Phys. Rev. Lett.},
  volume = {104},
  issue = {15},
  pages = {153601},
  numpages = {4},
  year = {2010},
  month = {Apr},
  publisher = {American Physical Society},
  doi = {10.1103/PhysRevLett.104.153601},
  url = {https://link.aps.org/doi/10.1103/PhysRevLett.104.153601}
}

@article{gao_observation_2015,
	title = {Observation of non-Hermitian degeneracies in a chaotic exciton-polariton billiard},
	volume = {526},
	issn = {0028-0836, 1476-4687},
	url = {https://www.nature.com/articles/nature15522},
	doi = {10.1038/nature15522},
	number = {7574},
	urldate = {2025-02-20},
	journal = {Nature},
	author = {Gao, T. and Estrecho, E. and Bliokh, K. Y. and Liew, T. C. H. and Fraser, M. D. and Brodbeck, S. and Kamp, M. and Schneider, C. and Höfling, S. and Yamamoto, Y. and Nori, F. and Kivshar, Y. S. and Truscott, A. G. and Dall, R. G. and Ostrovskaya, E. A.},
	month = oct,
	year = {2015},
	pages = {554--558},
}

@article{zhang_observation_2017,
	title = {Observation of the exceptional point in cavity magnon-polaritons},
	volume = {8},
	issn = {2041-1723},
	url = {https://www.nature.com/articles/s41467-017-01634-w},
	doi = {10.1038/s41467-017-01634-w},
	number = {1},
	urldate = {2025-02-20},
	journal = {Nat. Commun.},
	author = {Zhang, Dengke and Luo, Xiao-Qing and Wang, Yi-Pu and Li, Tie-Fu and You, J. Q.},
	month = nov,
	year = {2017},
	pages = {1368},
}

@article{PhysRevX.8.021066,
  title = {Dynamically Encircling Exceptional Points: In situ Control of Encircling Loops and the Role of the Starting Point},
  author = {Zhang, Xu-Lin and Wang, Shubo and Hou, Bo and Chan, C. T.},
  journal = {Phys. Rev. X},
  volume = {8},
  issue = {2},
  pages = {021066},
  numpages = {18},
  year = {2018},
  month = {Jun},
  publisher = {American Physical Society},
  doi = {10.1103/PhysRevX.8.021066},
  url = {https://link.aps.org/doi/10.1103/PhysRevX.8.021066}
}

@article{PhysRevLett.124.070402,
  title = {Tunable Nonreciprocal Quantum Transport through a Dissipative Aharonov-Bohm Ring in Ultracold Atoms},
  author = {Gou, Wei and Chen, Tao and Xie, Dizhou and Xiao, Teng and Deng, Tian-Shu and Gadway, Bryce and Yi, Wei and Yan, Bo},
  journal = {Phys. Rev. Lett.},
  volume = {124},
  issue = {7},
  pages = {070402},
  numpages = {6},
  year = {2020},
  month = {Feb},
  publisher = {American Physical Society},
  doi = {10.1103/PhysRevLett.124.070402},
  url = {https://link.aps.org/doi/10.1103/PhysRevLett.124.070402}
}

@article{PhysRevLett.126.170506,
  title = {Dynamically Encircling an Exceptional Point in a Real Quantum System},
  author = {Liu, Wenquan and Wu, Yang and Duan, Chang-Kui and Rong, Xing and Du, Jiangfeng},
  journal = {Phys. Rev. Lett.},
  volume = {126},
  issue = {17},
  pages = {170506},
  numpages = {5},
  year = {2021},
  month = {Apr},
  publisher = {American Physical Society},
  doi = {10.1103/PhysRevLett.126.170506},
  url = {https://link.aps.org/doi/10.1103/PhysRevLett.126.170506}
}

@article{doppler_dynamically_2016,
	title = {Dynamically encircling an exceptional point for asymmetric mode switching},
	volume = {537},
	issn = {0028-0836, 1476-4687},
	url = {https://www.nature.com/articles/nature18605},
	doi = {10.1038/nature18605},
	number = {7618},
	urldate = {2025-02-20},
	journal = {Nature},
	author = {Doppler, Jörg and Mailybaev, Alexei A. and Böhm, Julian and Kuhl, Ulrich and Girschik, Adrian and Libisch, Florian and Milburn, Thomas J. and Rabl, Peter and Moiseyev, Nimrod and Rotter, Stefan},
	month = sep,
	year = {2016},
	pages = {76--79},
}

@article{xu_topological_2016,
	title = {Topological energy transfer in an optomechanical system with exceptional points},
	volume = {537},
	issn = {0028-0836, 1476-4687},
	url = {https://www.nature.com/articles/nature18604},
	doi = {10.1038/nature18604},
	number = {7618},
	urldate = {2025-02-20},
	journal = {Nature},
	author = {Xu, H. and Mason, D. and Jiang, Luyao and Harris, J. G. E.},
	month = sep,
	year = {2016},
	pages = {80--83},
}

@article{yoon_time-asymmetric_2018,
	title = {Time-asymmetric loop around an exceptional point over the full optical communications band},
	volume = {562},
	issn = {0028-0836, 1476-4687},
	url = {https://www.nature.com/articles/s41586-018-0523-2},
	doi = {10.1038/s41586-018-0523-2},
	number = {7725},
	urldate = {2025-02-20},
	journal = {Nature},
	author = {Yoon, Jae Woong and Choi, Youngsun and Hahn, Choloong and Kim, Gunpyo and Song, Seok Ho and Yang, Ki-Yeon and Lee, Jeong Yub and Kim, Yongsung and Lee, Chang Seung and Shin, Jai Kwang and Lee, Hong-Seok and Berini, Pierre},
	month = oct,
	year = {2018},
	pages = {86--90},
}

@article{ren_chiral_2022,
	title = {Chiral control of quantum states in non-Hermitian spin–orbit-coupled fermions},
	volume = {18},
	issn = {1745-2473, 1745-2481},
	url = {https://www.nature.com/articles/s41567-021-01491-x},
	doi = {10.1038/s41567-021-01491-x},
	number = {4},
	urldate = {2025-02-20},
	journal = {Nat. Phys.},
	author = {Ren, Zejian and Liu, Dong and Zhao, Entong and He, Chengdong and Pak, Ka Kwan and Li, Jensen and Jo, Gyu-Boong},
	month = apr,
	year = {2022},
	pages = {385--389},
}

@article{PhysRevLett.131.260201,
  title = {Exceptional Entanglement Phenomena: Non-Hermiticity Meeting Nonclassicality},
  author = {Han, Pei-Rong and Wu, Fan and Huang, Xin-Jie and Wu, Huai-Zhi and Zou, Chang-Ling and Yi, Wei and Zhang, Mengzhen and Li, Hekang and Xu, Kai and Zheng, Dongning and Fan, Heng and Wen, Jianming and Yang, Zhen-Biao and Zheng, Shi-Biao},
  journal = {Phys. Rev. Lett.},
  volume = {131},
  issue = {26},
  pages = {260201},
  numpages = {7},
  year = {2023},
  month = {Dec},
  publisher = {American Physical Society},
  doi = {10.1103/PhysRevLett.131.260201},
  url = {https://link.aps.org/doi/10.1103/PhysRevLett.131.260201}
}

@article{bergholtz_exceptional_2021,
	title = {Exceptional topology of non-{Hermitian} systems},
	volume = {93},
	issn = {0034-6861, 1539-0756},
	url = {https://link.aps.org/doi/10.1103/RevModPhys.93.015005},
	doi = {10.1103/RevModPhys.93.015005},
	number = {1},
	urldate = {2025-02-20},
	journal = {Rev. Mod. Phys.},
	author = {Bergholtz, Emil J. and Budich, Jan Carl and Kunst, Flore K.},
	month = feb,
	year = {2021},
	pages = {015005},
}

@article{ding_non-hermitian_2022,
	title = {Non-Hermitian topology and exceptional-point geometries},
	volume = {4},
	issn = {2522-5820},
	url = {https://www.nature.com/articles/s42254-022-00516-5},
	doi = {10.1038/s42254-022-00516-5},
	number = {12},
	urldate = {2025-02-20},
	journal = {Nat. Rev. Phys.},
	author = {Ding, Kun and Fang, Chen and Ma, Guancong},
	month = oct,
	year = {2022},
	pages = {745--760},
}

@article{chen_exceptional_2017,
	title = {Exceptional points enhance sensing in an optical microcavity},
	volume = {548},
	issn = {0028-0836, 1476-4687},
	url = {https://www.nature.com/articles/nature23281},
	doi = {10.1038/nature23281},
	number = {7666},
	urldate = {2025-02-20},
	journal = {Nature},
	author = {Chen, Weijian and Kaya Özdemir, Sahin and Zhao, Guangming and Wiersig, Jan and Yang, Lan},
	month = aug,
	year = {2017},
	pages = {192--196},
}

@article{hodaei_enhanced_2017,
	title = {Enhanced sensitivity at higher-order exceptional points},
	volume = {548},
	issn = {0028-0836, 1476-4687},
	url = {https://www.nature.com/articles/nature23280},
	doi = {10.1038/nature23280},
	number = {7666},
	urldate = {2025-02-20},
	journal = {Nature},
	author = {Hodaei, Hossein and Hassan, Absar U. and Wittek, Steffen and Garcia-Gracia, Hipolito and El-Ganainy, Ramy and Christodoulides, Demetrios N. and Khajavikhan, Mercedeh},
	month = aug,
	year = {2017},
	pages = {187--191},
}

@article{miri_exceptional_2019,
	title = {Exceptional points in optics and photonics},
	volume = {363},
	issn = {0036-8075, 1095-9203},
	url = {https://www.science.org/doi/10.1126/science.aar7709},
	doi = {10.1126/science.aar7709},
	number = {6422},
	urldate = {2025-02-20},
	journal = {Science},
	author = {Miri, Mohammad-Ali and Alù, Andrea},
	month = jan,
	year = {2019},
	pages = {eaar7709},
}

@article{ozdemir_paritytime_2019,
	title = {Parity–time symmetry and exceptional points in photonics},
	volume = {18},
	issn = {1476-1122, 1476-4660},
	url = {https://www.nature.com/articles/s41563-019-0304-9},
	doi = {10.1038/s41563-019-0304-9},
	number = {8},
	urldate = {2025-02-20},
	journal = {Nat. Mater.},
	author = {Özdemir, S. K. and Rotter, S. and Nori, F. and Yang, L.},
	month = aug,
	year = {2019},
	pages = {783--798},
}

@article{PhysRevLett.118.045701,
  title = {Weyl Exceptional Rings in a Three-Dimensional Dissipative Cold Atomic Gas},
  author = {Xu, Yong and Wang, Sheng-Tao and Duan, L.-M.},
  journal = {Phys. Rev. Lett.},
  volume = {118},
  issue = {4},
  pages = {045701},
  numpages = {5},
  year = {2017},
  month = {Jan},
  publisher = {American Physical Society},
  doi = {10.1103/PhysRevLett.118.045701},
  url = {https://link.aps.org/doi/10.1103/PhysRevLett.118.045701}
}

@article{PhysRevB.99.121101,
  title = {Symmetry-protected exceptional rings in two-dimensional correlated systems with chiral symmetry},
  author = {Yoshida, Tsuneya and Peters, Robert and Kawakami, Norio and Hatsugai, Yasuhiro},
  journal = {Phys. Rev. B},
  volume = {99},
  issue = {12},
  pages = {121101},
  numpages = {5},
  year = {2019},
  month = {Mar},
  publisher = {American Physical Society},
  doi = {10.1103/PhysRevB.99.121101},
  url = {https://link.aps.org/doi/10.1103/PhysRevB.99.121101}
}

@article{PhysRevLett.127.196801,
  title = {Higher-Order Weyl-Exceptional-Ring Semimetals},
  author = {Liu, Tao and He, James Jun and Yang, Zhongmin and Nori, Franco},
  journal = {Phys. Rev. Lett.},
  volume = {127},
  issue = {19},
  pages = {196801},
  numpages = {9},
  year = {2021},
  month = {Nov},
  publisher = {American Physical Society},
  doi = {10.1103/PhysRevLett.127.196801},
  url = {https://link.aps.org/doi/10.1103/PhysRevLett.127.196801}
}

@article{PhysRevB.104.L161117,
  title = {Non-Hermitian higher-order Weyl semimetals},
  author = {Ghorashi, Sayed Ali Akbar and Li, Tianhe and Sato, Masatoshi},
  journal = {Phys. Rev. B},
  volume = {104},
  issue = {16},
  pages = {L161117},
  numpages = {7},
  year = {2021},
  month = {Oct},
  publisher = {American Physical Society},
  doi = {10.1103/PhysRevB.104.L161117},
  url = {https://link.aps.org/doi/10.1103/PhysRevB.104.L161117}
}

@article{PhysRevResearch.2.043268,
  title = {Weyl points and exceptional rings with polaritons in bulk semiconductors},
  author = {Mc Guinness, R. L. and Eastham, P. R.},
  journal = {Phys. Rev. Res.},
  volume = {2},
  issue = {4},
  pages = {043268},
  numpages = {10},
  year = {2020},
  month = {Nov},
  publisher = {American Physical Society},
  doi = {10.1103/PhysRevResearch.2.043268},
  url = {https://link.aps.org/doi/10.1103/PhysRevResearch.2.043268}
}

@article{PhysRevB.100.245205,
  title = {Disorder-induced exceptional and hybrid point rings in Weyl/Dirac semimetals},
  author = {Matsushita, Taiki and Nagai, Yuki and Fujimoto, Satoshi},
  journal = {Phys. Rev. B},
  volume = {100},
  issue = {24},
  pages = {245205},
  numpages = {9},
  year = {2019},
  month = {Dec},
  publisher = {American Physical Society},
  doi = {10.1103/PhysRevB.100.245205},
  url = {https://link.aps.org/doi/10.1103/PhysRevB.100.245205}
}

@article{PhysRevLett.129.084301,
  title = {Experimental Realization of Weyl Exceptional Rings in a Synthetic Three-Dimensional Non-Hermitian Phononic Crystal},
  author = {Liu, Jing-jing and Li, Zheng-wei and Chen, Ze-Guo and Tang, Weiyuan and Chen, An and Liang, Bin and Ma, Guancong and Cheng, Jian-Chun},
  journal = {Phys. Rev. Lett.},
  volume = {129},
  issue = {8},
  pages = {084301},
  numpages = {7},
  year = {2022},
  month = {Aug},
  publisher = {American Physical Society},
  doi = {10.1103/PhysRevLett.129.084301},
  url = {https://link.aps.org/doi/10.1103/PhysRevLett.129.084301}
}

@article{zhen_spawning_2015,
	title = {Spawning rings of exceptional points out of Dirac cones},
	volume = {525},
	issn = {0028-0836, 1476-4687},
	url = {https://www.nature.com/articles/nature14889},
	doi = {10.1038/nature14889},
	number = {7569},
	urldate = {2025-02-20},
	journal = {Nature},
	author = {Zhen, Bo and Hsu, Chia Wei and Igarashi, Yuichi and Lu, Ling and Kaminer, Ido and Pick, Adi and Chua, Song-Liang and Joannopoulos, John D. and Soljačić, Marin},
	month = sep,
	year = {2015},
	pages = {354--358},
}

@article{cerjan_experimental_2019,
	title = {Experimental realization of a {Weyl} exceptional ring},
	volume = {13},
	issn = {1749-4885, 1749-4893},
	url = {https://www.nature.com/articles/s41566-019-0453-z},
	doi = {10.1038/s41566-019-0453-z},
	number = {9},
	urldate = {2025-02-20},
	journal = {Nat. Photonics},
	author = {Cerjan, Alexander and Huang, Sheng and Wang, Mohan and Chen, Kevin P. and Chong, Yidong and Rechtsman, Mikael C.},
	month = sep,
	year = {2019},
	pages = {623--628},
}

@article{tang_realization_2023,
	title = {Realization and topological properties of third-order exceptional lines embedded in exceptional surfaces},
	volume = {14},
	issn = {2041-1723},
	url = {https://www.nature.com/articles/s41467-023-42414-z},
	doi = {10.1038/s41467-023-42414-z},
	number = {1},
	urldate = {2025-02-20},
	journal = {Nat. Commun.},
	author = {Tang, Weiyuan and Ding, Kun and Ma, Guancong},
	month = oct,
	year = {2023},
	pages = {6660},
}

@article{zhou_exceptional_2019,
	title = {Exceptional surfaces in {PT}-symmetric non-{Hermitian} photonic systems},
	volume = {6},
	issn = {2334-2536},
	url = {https://opg.optica.org/abstract.cfm?URI=optica-6-2-190},
	doi = {10.1364/OPTICA.6.000190},
	number = {2},
	urldate = {2025-02-20},
	journal = {Optica},
	author = {Zhou, Hengyun and Lee, Jong Yeon and Liu, Shang and Zhen, Bo},
	month = feb,
	year = {2019},
	pages = {190},
}

@article{PhysRevLett.123.237202,
  title = {Experimental Observation of an Exceptional Surface in Synthetic Dimensions with Magnon Polaritons},
  author = {Zhang, Xufeng and Ding, Kun and Zhou, Xianjing and Xu, Jing and Jin, Dafei},
  journal = {Phys. Rev. Lett.},
  volume = {123},
  issue = {23},
  pages = {237202},
  numpages = {5},
  year = {2019},
  month = {Dec},
  publisher = {American Physical Society},
  doi = {10.1103/PhysRevLett.123.237202},
  url = {https://link.aps.org/doi/10.1103/PhysRevLett.123.237202}
}

@book{sakurai_modern_2011,
	address = {Boston, Mass.},
	edition = {2. ed., international ed},
	title = {Modern quantum mechanics},
	isbn = {9780805382914 9781292024103},
	publisher = {Addison-Wesley, Pearson},
	author = {Sakurai, Jun J. and Napolitano, Jim},
	year = {2011},
}

@article{uhlmann_parallel_1986,
	title = {Parallel transport and “quantum holonomy” along density operators},
	volume = {24},
	copyright = {https://www.elsevier.com/tdm/userlicense/1.0/},
	issn = {00344877},
	url = {https://linkinghub.elsevier.com/retrieve/pii/0034487786900558},
	doi = {10.1016/0034-4877(86)90055-8},
	number = {2},
	urldate = {2025-02-20},
	journal = {Rep. Math. Phys.},
	author = {Uhlmann, Armin},
	month = oct,
	year = {1986},
	pages = {229--240},
}

@article{uhlmann_gauge_1991,
	title = {A gauge field governing parallel transport along mixed states},
	volume = {21},
	copyright = {http://www.springer.com/tdm},
	issn = {0377-9017, 1573-0530},
	url = {http://link.springer.com/10.1007/BF00420373},
	doi = {10.1007/BF00420373},
	number = {3},
	urldate = {2025-02-20},
	journal = {Lett. Math. Phys.},
	author = {Uhlmann, Armin},
	month = mar,
	year = {1991},
	pages = {229--236},
}

@article{uhlmann_density_1993,
	title = {Density operators as an arena for differential geometry},
	volume = {33},
	copyright = {https://www.elsevier.com/tdm/userlicense/1.0/},
	issn = {00344877},
	url = {https://linkinghub.elsevier.com/retrieve/pii/003448779390060R},
	doi = {10.1016/0034-4877(93)90060-R},
	number = {1-2},
	urldate = {2025-02-20},
	journal = {Rep. Math. Phys.},
	author = {Uhlmann, Armin},
	month = aug,
	year = {1993},
	pages = {253--263},
}

@article{PhysRevLett.91.090405,
  title = {Mixed State Geometric Phases, Entangled Systems, and Local Unitary Transformations},
  author = {Ericsson, Marie and Pati, Arun K. and Sj\"oqvist, Erik and Br\"annlund, Johan and Oi, Daniel K. L.},
  journal = {Phys. Rev. Lett.},
  volume = {91},
  issue = {9},
  pages = {090405},
  numpages = {4},
  year = {2003},
  month = {Aug},
  publisher = {American Physical Society},
  doi = {10.1103/PhysRevLett.91.090405},
  url = {https://link.aps.org/doi/10.1103/PhysRevLett.91.090405}
}

@article{PhysRevA.75.032106,
  title = {Operational approach to the Uhlmann holonomy},
  author = {\AA{}berg, Johan and Kult, David and Sj\"oqvist, Erik and Oi, Daniel K. L.},
  journal = {Phys. Rev. A},
  volume = {75},
  issue = {3},
  pages = {032106},
  numpages = {7},
  year = {2007},
  month = {Mar},
  publisher = {American Physical Society},
  doi = {10.1103/PhysRevA.75.032106},
  url = {https://link.aps.org/doi/10.1103/PhysRevA.75.032106}
}

@article{zhu_experimental_2011,
	title = {Experimental demonstration of a unified framework for mixed-state geometric phases},
	volume = {94},
	issn = {0295-5075, 1286-4854},
	url = {https://iopscience.iop.org/article/10.1209/0295-5075/94/20007},
	doi = {10.1209/0295-5075/94/20007},
	number = {2},
	urldate = {2025-02-20},
	journal = {Europhys. Lett.},
	author = {Zhu, J. and Shi, M. and Vedral, V. and Peng, X. and Suter, D. and Du, J.},
	month = apr,
	year = {2011},
	pages = {20007},
}

@article{PhysRevB.91.165140,
  title = {Topology of density matrices},
  author = {Budich, Jan Carl and Diehl, Sebastian},
  journal = {Phys. Rev. B},
  volume = {91},
  issue = {16},
  pages = {165140},
  numpages = {8},
  year = {2015},
  month = {Apr},
  publisher = {American Physical Society},
  doi = {10.1103/PhysRevB.91.165140},
  url = {https://link.aps.org/doi/10.1103/PhysRevB.91.165140}
}

@article{andersson_geometric_2016,
	title = {Geometric phases for mixed states of the Kitaev chain},
	volume = {374},
	issn = {1364-503X, 1471-2962},
	url = {https://royalsocietypublishing.org/doi/10.1098/rsta.2015.0231},
	doi = {10.1098/rsta.2015.0231},
	number = {2068},
	urldate = {2025-02-20},
	journal = {Philos. Trans. R. Soc. A},
	author = {Andersson, Ole and Bengtsson, Ingemar and Ericsson, Marie and Sjöqvist, Erik},
	month = may,
	year = {2016},
	pages = {20150231},
}

@article{mera_boltzmanngibbs_2017,
	title = {Boltzmann–{Gibbs} states in topological quantum walks and associated many-body systems: fidelity and {Uhlmann} parallel transport analysis of phase transitions},
	volume = {50},
	issn = {1751-8113, 1751-8121},
	shorttitle = {Boltzmann–{Gibbs} states in topological quantum walks and associated many-body systems},
	url = {https://iopscience.iop.org/article/10.1088/1751-8121/aa820e},
	doi = {10.1088/1751-8121/aa820e},
	number = {36},
	urldate = {2025-02-20},
	journal = {J. Phys. A: Math. Theor.},
	author = {Mera, Bruno and Vlachou, Chrysoula and Paunković, Nikola and Vieira, Vítor R},
	month = sep,
	year = {2017},
	pages = {365302},
}

@article{PhysRevLett.119.015702,
  title = {Uhlmann Connection in Fermionic Systems Undergoing Phase Transitions},
  author = {Mera, Bruno and Vlachou, Chrysoula and Paunković, Nikola and Vieira, Vítor R},
  journal = {Phys. Rev. Lett.},
  volume = {119},
  issue = {1},
  pages = {015702},
  numpages = {6},
  year = {2017},
  month = {Jul},
  publisher = {American Physical Society},
  doi = {10.1103/PhysRevLett.119.015702},
  url = {https://link.aps.org/doi/10.1103/PhysRevLett.119.015702}
}

@article{viyuela_observation_2018,
	title = {Observation of topological Uhlmann phases with superconducting qubits},
	volume = {4},
	issn = {2056-6387},
	url = {https://www.nature.com/articles/s41534-017-0056-9},
	doi = {10.1038/s41534-017-0056-9},
	number = {1},
	urldate = {2025-02-20},
	journal = {npj Quantum Inf.},
	author = {Viyuela, O. and Rivas, A. and Gasparinetti, S. and Wallraff, A. and Filipp, S. and Martin-Delgado, M. A.},
	month = feb,
	year = {2018},
	pages = {10},
}

@article{PhysRevLett.112.130401,
  title = {Uhlmann Phase as a Topological Measure for One-Dimensional Fermion Systems},
  author = {Viyuela, O. and Rivas, A. and Martin-Delgado, M. A.},
  journal = {Phys. Rev. Lett.},
  volume = {112},
  issue = {13},
  pages = {130401},
  numpages = {5},
  year = {2014},
  month = {Apr},
  publisher = {American Physical Society},
  doi = {10.1103/PhysRevLett.112.130401},
  url = {https://link.aps.org/doi/10.1103/PhysRevLett.112.130401}
}

@article{PhysRevB.106.024310,
  title = {Uhlmann holonomy against Lindblad dynamics of topological systems at finite temperatures},
  author = {He, Yan and Chien, Chih-Chun},
  journal = {Phys. Rev. B},
  volume = {106},
  issue = {2},
  pages = {024310},
  numpages = {10},
  year = {2022},
  month = {Jul},
  publisher = {American Physical Society},
  doi = {10.1103/PhysRevB.106.024310},
  url = {https://link.aps.org/doi/10.1103/PhysRevB.106.024310}
}

@article{PhysRevB.97.235141,
  title = {Thermal Uhlmann-Chern number from the Uhlmann connection for extracting topological properties of mixed states},
  author = {He, Yan and Guo, Hao and Chien, Chih-Chun},
  journal = {Phys. Rev. B},
  volume = {97},
  issue = {23},
  pages = {235141},
  numpages = {7},
  year = {2018},
  month = {Jun},
  publisher = {American Physical Society},
  doi = {10.1103/PhysRevB.97.235141},
  url = {https://link.aps.org/doi/10.1103/PhysRevB.97.235141}
}

@article{kartik_mixed_2023,
	title = {Mixed state behavior of Hermitian and non-Hermitian topological models with extended couplings},
	volume = {13},
	issn = {2045-2322},
	url = {https://www.nature.com/articles/s41598-023-33449-9},
	doi = {10.1038/s41598-023-33449-9},
	number = {1},
	urldate = {2025-02-20},
	journal = {Sci. Rep.},
	author = {Kartik, Y. R. and Sarkar, Sujit},
	month = apr,
	year = {2023},
	pages = {6431},
}

@article{PhysRevB.105.085418,
  title = {Proxy ensemble geometric phase and proxy index of time-reversal invariant topological insulators at finite temperatures},
  author = {Pi, Aixin and Zhang, Ye and He, Yan and Chien, Chih-Chun},
  journal = {Phys. Rev. B},
  volume = {105},
  issue = {8},
  pages = {085418},
  numpages = {10},
  year = {2022},
  month = {Feb},
  publisher = {American Physical Society},
  doi = {10.1103/PhysRevB.105.085418},
  url = {https://link.aps.org/doi/10.1103/PhysRevB.105.085418}
}

@article{PhysRevResearch.5.023004,
  title = {Topological phase transitions at finite temperature},
  author = {Molignini, Paolo and Cooper, Nigel R.},
  journal = {Phys. Rev. Res.},
  volume = {5},
  issue = {2},
  pages = {023004},
  numpages = {15},
  year = {2023},
  month = {Apr},
  publisher = {American Physical Society},
  doi = {10.1103/PhysRevResearch.5.023004},
  url = {https://link.aps.org/doi/10.1103/PhysRevResearch.5.023004}
}

@article{PhysRevLett.85.2845,
  title = {Geometric Phases for Mixed States in Interferometry},
  author = {Sj\"oqvist, Erik and Pati, Arun K. and Ekert, Artur and Anandan, Jeeva S. and Ericsson, Marie and Oi, Daniel K. L. and Vedral, Vlatko},
  journal = {Phys. Rev. Lett.},
  volume = {85},
  issue = {14},
  pages = {2845--2849},
  numpages = {0},
  year = {2000},
  month = {Oct},
  publisher = {American Physical Society},
  doi = {10.1103/PhysRevLett.85.2845},
  url = {https://link.aps.org/doi/10.1103/PhysRevLett.85.2845}
}

@article{PhysRevX.8.011035,
  title = {Probing the Topology of Density Matrices},
  author = {Bardyn, Charles-Edouard and Wawer, Lukas and Altland, Alexander and Fleischhauer, Michael and Diehl, Sebastian},
  journal = {Phys. Rev. X},
  volume = {8},
  issue = {1},
  pages = {011035},
  numpages = {21},
  year = {2018},
  month = {Feb},
  publisher = {American Physical Society},
  doi = {10.1103/PhysRevX.8.011035},
  url = {https://link.aps.org/doi/10.1103/PhysRevX.8.011035}
}
\end{document}